\documentclass  [
  paper    = a4,
  BCOR     = 11pt
]{paper}
\pdfoutput=1

\usepackage{packages}
\usepackage{commands}
\usepackage{jcappub}

\notoc
\title{Kinetic Field Theory: Perturbation theory beyond first order}

\author[a]{C. Pixius,}
\author[a]{{S}. {C}elik}
\author[a]{and M. Bartelmann}

\affiliation[a]{Heidelberg University, Institut f\"ur Theoretische Physik, Philosophenweg 12, D-69120 Heidelberg, Germany}

\emailAdd{pixius@thphys.uni-heidelberg.de}
\emailAdd{safakcelik12@gmail.com}
\emailAdd{bartelmann@uni-heidelberg.de}

\abstract{
We present recent improvements in the perturbative treatment of particle interactions in Kinetic Field Theory (KFT) for inertial Zel'dovich trajectories.
KFT has been developed for the systematic analytical calculation of non-linear cosmic structure formation on the basis of microscopic phase-space dynamics. 
We improve upon the existing treatment of the interaction operator by deriving a more rigorous treatment of phase-space trajectories of particles in an expanding universe.
We then show how these results can be applied to KFT perturbation theory by calculating corrections to the late-time dark matter power spectrum at second order in the interaction operator. 
We find that the modified treatment of interactions w.r.t.\ inertial Zel'dovich trajectories improves the agreement of KFT with simulation results on intermediate scales compared to earlier results.
Additionally, we illustrate that including particle interactions up to second order leads to a systematic improvement of the non-linear power spectrum compared to the first-order result.
}


\begin{document}

\maketitle

\section{Introduction}
\label{sec:introduction}
The analytical description of large-scale cosmic structure formation with cold dark matter is a long standing challenge in modern cosmology.
A proper understanding of the dynamics of dark matter under the influence of gravity could help to constrain the cosmological model and explore the nature of gravity \cite{AMICO, ref_Euclid_validation, 2011ARA&A..49..409A_Allen}, and has therefore been the focus of intense research in the past two decades.
On the one hand, traditional methods such as Eulerian (standard) perturbation theory (SPT) and Lagrangian perturbation theory (LPT) are very successful on intermediate to large-scales; see \cite{BERNARDEAU20021} for a comprehensive review.
The seminal work of Crocce and Scoccimarro \cite{2006PhRvD..73f3519C, 2006PhRvD..73f3520C} has led to the development of a diverse panoply of methods building upon the foundations of SPT and LPT but going beyond loop expansion schemes \cite{Matarrese_2007, 1607.03453_floerchingerRG, Erschfeld:2021ocz}.

Nevertheless, an extension to highly non-linear scales has not yet been achieved.
For Eulerian schemes the assumption of a unique velocity field is a fundamental obstacle and causes the breakdown of the theory once streams of dark matter particles cross \cite{PhysRevD.75.021302_ashford}.
In its standard form LPT suffers from similar drawbacks, although there have been valid efforts to extend it to account for crossing particle streams \cite{PhysRevD.97.023508_McDonald_Vlah, PhysRevD.87.083522_Valageas}.

On the other hand $N$-body simulations show impressive results and are nowadays the main tool to explore structure growth on highly non-linear scales.
But despite their remarkable success, simulations have a couple of drawbacks; 
they yield little insight into the actual processes underlying large-scale structure formation and they are computationally costly.
With a growing interest in different models of gravity, testing cosmic structure formation for a large number of physical models will be a challenge in the future that simulations can hardly meet by themselves.

Kinetic Field Theory (KFT), which has recently been developed as an alternative analytical method to describe the large-scale cosmic structure formation takes ideas from both simulations and previous analytical methods.
Building upon earlier developments in statistical field theory \cite{1973PhRvA...8..423M, 2012JSP...149..643D, 2013JSP...152..159D}, KFT describes the evolution of cosmic structures on the basis of the motion of classical particles in phase-space.
Since particle trajectories do not cross there, this allows for a well-defined treatment of structure formation beyond stream crossing and can be regarded as an analytical analogue to $N$-body simulations.
In \cite{2016NJPh...18d3020B} a first application of KFT Perturbation Theory (KPT) was applied to calculate the non-linear density fluctuation power spectrum in cosmology.
The idea of KPT is radically different from other cosmological perturbation theories in that we perturb microscopic particle trajectories instead of macroscopic fields, such as density or velocity fields.
{Although we will focus exclusively on non-linear structure formation in dark matter, adding baryons into the picture is possible as well in KFT. 
This is most easily done in the framework of a macroscopic reformulation of the theory presented in \cite{2018arXiv180906942L}.
}

In this paper we want to show recent developments in KPT by introducing improvements upon the results for first-order perturbations of \cite{2016NJPh...18d3020B} and deriving expressions for second-order perturbations.
We will start with a short introduction of the main ideas of KFT in \autoref{sec:generating functional}, where we derive the generating functional which is the key object of the theory.
Additionally, we will discuss the effect of the interaction operator on the free generating functional.
In \autoref{sec:particle trajectories} we introduce an expression for the phase-space trajectories of classical particles in an expanding universe and discuss how this expression needs to be modified to recover linear growth of the density fluctuation power spectrum on very large scales.
A thorough treatment of perturbations of particle trajectories up to second order for a power spectrum is given in \autoref{sec:perturbation theory}. 
We employ the diagrammatic approach of \cite{2017NJPh...19h3001B} and link it to the analytical formalism that was presented in \cite{2015PhRvE..91f2120V}.
The results for the non-linear power spectrum at first- and second-order KPT are presented and discussed in \autoref{sec:numerical results}.

\subsection{Notation}
Since the fundamental fields in KFT are the $6N$ phase-space coordinates of the classical particles of the system under consideration we want to introduce a notation that allows us to write them in a compact form.
For each individual particle we collect the phase-space coordinates in a single vector $\vec{x}_j = (\vecq_j, \vecp_j)^{\mathrm{T}}$, with the particle index $j = 1, \dotsc , N$.
We then define
\begin{equation}
    \tensorx(t) \defeq \sum_{j=1}^N \Vec{x}_j(t) \otimes \Vec{e}_j 
    = \sum_{j=1}^N 
    \begin{pmatrix}
    \vecq_j(t) \\ \vecp_j(t)
    \end{pmatrix}
    \otimes \Vec{e}_j,
\end{equation}
where the only non-vanishing entry of $\Vec{e}_j$ is a 1 at position $j$. 
This bundles the phase-space coordinates of all the particles of the system into one single tensorial object.
We define a scalar product for such tensors $\mathbf{A}$ and $\mathbf{B}$ by 
\begin{equation*}
    \mathbf{A} \cdot \mathbf{B} \defeq 
    \sum_{j=1}^N \vec{A}_j \cdot \vec{B}_j.
\end{equation*}
If the tensors are additionally time-dependent we extend the definition of the scalar product to include a time integral
\begin{equation*}
    \mathbf{A} \cdot \mathbf{B} \defeq 
    \sum_{j=1}^N \int \dd t \ \vec{A}_j(t) \cdot \vec{B}_j(t).
\end{equation*}
Fields that carry space-time coordinates as their arguments also arise in a natural way in KFT, most notably when we consider interactions and they are usually expressed in Fourier space, $f(\vec{k},t)$.
To allow for a compact notation of the arguments of such fields we introduce field labels $r$ such that $f(\pm r) = f(\pm \veck_r, t_r)$.
A Fourier space-time integral over such a function can then be written as
\begin{align*}
    \int \dd 1 f(1) \defeq \int \frac{\dd^3 k_1}{(2\pi)^3} \int \dd t_1 \ f(\veck_1, t_1).
\end{align*}
Finally, for spatial integrals in configuration- and Fourier space we use the shorthand notations
\begin{align*}
    \int_q \defeq \int \dd^3 q,
    \qquad \qquad
    \int_k \defeq \int \frac{\dd^3 k}{(2\pi)^3}.
\end{align*}

\section{Generating functional}
\label{sec:generating functional}

KFT allows calculating statistical properties of classical particle ensembles that may be far from equilibrium.
Cosmic structure formation is an example of such an ensemble that is of particular interest, but applications go beyond.
In any case we assume that the system under consideration consists of a larger number $N$ of classical point particles, interacting via a two-particle interaction potential $v(\vecq_i - \vecq_j)$. Here $\vecq_i$ and $\vecq_j$ denote the respective positions of particles $i$ and $j$.
We assume that we have statistical knowledge about the initial conditions of the system, which can be translated into an initial distribution of the particles' phase-space coordinates, e.g.\ via Poisson sampling.
%The initial probability distribution of phase-space coordinates can be derived from the distribution of the initial macroscopic quantities, e.g.\ via Poisson sampling.
One should keep in mind, however, that the properties of the distribution of phase-space coordinates are in general very different from those of the underlying macroscopic fields.
In cosmology the initial density and velocity fields are commonly assumed to be Gaussian or very close to it. 
The resulting distribution for the phase-space coordinates of the microscopic particles of the system, on the other hand, is non-Gaussian even initially.
Macroscopic features of the system, such as density or velocity correlation functions can be formally determined using the canonical partition function.
Since particles move along classical trajectories, there is a unique initial phase-space configuration $\tensorxin$ that gives rise to the final configuration $\tensorx^{(\mathrm{f})}$.
We can use this fact to write the canonical partition function in terms of an integral over the initial probability distribution and a transition probability between initial and final states
\begin{align}\label{eq:canonical partition funct}
	Z &= \int \dd \tensorx^{(\mathrm{f})} \int \dd \tensorxin P(\tensorxin) 
	\int_{\tensorx(t_\mathrm{i}) = \tensorxin}^{\tensorx(t_\mathrm{f}) = \tensorx^{(\mathrm{f})}} \DD \tensorx (t) \
	\delta_D\big[\tensorx(t) - \tilde{\tensorx}(t;\tensorxin) \big] \nonumber
	\\
	& = \int \dd \tensorxin P(\tensorxin) 
	\int_{\mathrm{i}} \DD \tensorx (t) \
	\delta_D\big[\dot{\tensorx} - \mathcal{J} \nabla_\tensorx \HH \big].
\end{align}
In the first line of \eqref{eq:canonical partition funct}, $\tilde{\tensorx}(t; \tensorxin)$ denotes the classical trajectory in the $6N$-dimensional phase-space (which is unique). 
Essentially, we sum over all possible phase-space paths that connect the initial conditions with the state at time $t$ and give each path a weight.
This is in complete analogy to Feynman's approach to evaluating transition matrix elements in quantum mechanics.
The key difference is that there is only one single path that is allowed for a classical system, which results in the fact that the weight we attribute to each path is represented by a Dirac delta distribution.
In line 2 the argument has been replaced by the equations of motion of a system with Hamiltonian $\HH$. Here $\mathcal{J} = J \otimes \mathcal{I}_N$, where $J$ is the symplectic matrix and $\mathcal{I}_N$ the $N$-dimensional unit matrix.
The step leading from line 1 to line 2 involves a Jacobian that can be shown to be a constant \cite{PhysRevD.40.3363_Gozzi} which is then reabsorbed into the integration measure.

We can represent the functional Dirac distribution as a functional Fourier integral by introducing a conjugated field $\tensorchi(t)$, and promote the canonical partition function to a generating functional by endowing it with two generator fields $\tensorJ(t)$ and $\tensorK(t)$, yielding
\begin{align}\label{eq:genFunc1}
    Z[\tensorJ, \tensorK] = 
    \int \dgamma
	\int_{\mathrm{i}} \DD \tensorx (t)
	\int \DD \tensorchi(t)\
	\exp\big\{\ii \tensorchi \cdot (\dot{\tensorx} - \mathcal{J}\nabla_\tensorx \HH_0) 
	-\ii \tensorchi \cdot \mathcal{J}\nabla_\tensorx \HH_\mathcal{I}\big\} \ee^{\ii \tensorJ \cdot \tensorx} \ee^{\ii \tensorK \cdot \tensorchi}.
\end{align}
Note that we have abbreviated the initial probability measure by $\dgamma = \dd \tensorxin P(\tensorxin)$.
Additionally, we have split the Hamiltonian into a contribution $\HH_0$ giving rise to inertial motion, as well as a contribution due to particle interactions $\HH_\mathcal{I}$. This splitting is completely arbitrary, and allows us to choose inertial trajectories that are most suitable for the system at hand. 
The argument of the exponential function in \eqref{eq:genFunc1} is the MSR action
\begin{equation}\label{eq:action}
    \ii S = \ii S_0 + \ii S_\mathcal{I} \defeq \ii \tensorchi \cdot (\dot{\tensorx} - \mathcal{J}\nabla_\tensorx \HH_0) 
	-\ii \tensorchi \cdot \mathcal{J}\nabla_\tensorx \HH_\mathcal{I},
\end{equation}
which generates the equations of motion via $\frac{\delta S}{\delta \tensorchi} = 0$ \cite{1973PhRvA...8..423M}.
To allow for a perturbative treatment of the interacting part of the action $S_\mathcal{I}$ we promote it to an operator $\hat{S}_\mathcal{I}$ by replacing
\begin{equation}\label{eq:transform to operators}
    \tensorchi(t) \rightarrow \hat{\tensorchi}(t) = \frac{\delta}{\ii \delta \tensorK (t)},
    \qquad  \qquad
    \tensorx(t) \rightarrow \hat{\tensorx}(t) = \frac{\delta}{\ii\delta \tensorJ (t)},
\end{equation}
and pull it in front of the integrals. As a result the generating functional splits into a free part and an interaction operator:
\begin{align}\label{eq:genFunc2}
	Z[\tensorJ, \tensorK] 
	&= \exp \big( \ii \hat{S}_\mathcal{I} \big)
	\int \dgamma
	\int_{\mathrm{i}} \DD \tensorx (t)
	\int \DD \tensorchi(t)\
	\ee^{\ii S_0} \ee^{\ii \tensorJ \cdot \tensorx} \ee^{\ii \tensorK \cdot \tensorchi}\nonumber
	\\
	&\eqdef
	\exp \big( \ii\hat{S}_\mathcal{I} \big) 
	Z_0[\tensorJ,\tensorK].
\end{align}

\begin{comment}
By solving the remaining path integrals over $\tensorx(t)$ and $\tensorchi(t)$ we have enforced the condition that particles follow classical trajectories with an inhomogeneity $\tensorK(t)$:
\begin{equation}
	\tilde{\tensorx}[\tensorK](t) = \boldsymbol{\mathcal{G}}(t,0) \, \tensorxin - \int_0^t \dd\bar{t} \ \boldsymbol{\mathcal{G}}(t,\bar{t}) \,\tensorK(\bar{t}).
\end{equation}
Here $\boldsymbol{\mathcal{G}}(t,\bar{t})$ is the Green's function that solves the free equation of motion $\frac{\delta S_0}{\delta \tensorchi} = 0$.
\end{comment}

%%%%%%%%%%%%%%%%%%%%%%%%%%%%%%%%%%%%%%%%%%%%%%%%%%%%%%%
%%%%%%%%%%%%%%%%%%%%%%%%%%%%%%%%%%%%%%%%%%%%%%%%%%%%%%%
\subsection{Density and response field operators}
Before addressing the evaluation of the interaction operator $\hat{S}_\mathcal{I}$ we need to introduce an expression for the particle density field $\rho(\vecq,t)$ in terms of the phase-space coordinates $\tensorx(t)$.
In KFT treat the density field as a sum of classical point particles
\begin{align}\label{eq:particle density}
    \rho(\vecq,t) = \sum_{j=1}^N \delta_D (\vecq - \vecq_j(t)),
\end{align}
where $\vecq_j(t)$ denotes the real-space position of particle $j$ at time $t$.
Going to Fourier space we find
\begin{align*}
    \rho(\veck,t) = \sum_{j=1}^N \ee^{- \ii \veck \cdot \vecq_j(t)} = \sum_{j=1}^N\rho_j(\veck,t),
\end{align*}
which can be promoted to an operator $\hat{\rho} (\veck,t)$ by applying the rules \eqref{eq:transform to operators}.
Assuming that interactions can be represented by a collective potential $V(\vecq,t)$, the interacting part of the MSR action \eqref{eq:action} in operator form is given by
\begin{equation}\label{eq:interaction operator 1}
    \hat{S}_\mathcal{I}
    = \hat{\tensorchi}_p \cdot {\boldsymbol{\nabla} V(\hat{\tensorq})}.
\end{equation}
We are interested in the case of the collective potential being the result of two-body interactions, such that
\begin{align}\label{eq:collectivePotential}
    V(\vecq,t) = \sum_{j=1}^N v(\vecq - \vecq_j(t)) = \int_y \rho(\vecy,t) v(\vecq - \vecy,t),
\end{align}
where the second equality follows from the definition \eqref{eq:particle density} of the density field.
To evaluate the gradient of the collective potential we write
\begin{align}\label{eq:potentialGradient}
    {\boldsymbol{\nabla} V({\hat{\tensorq}})}
    &= \sum_{j=1}^N \vnabla V(\vecq,t) \Big|_{\vecq = \hat{\vecq}_j(t)} \otimes \vec{e}_j \nonumber
    \\
    &=\int_q \vnabla V(\vecq, t) \sum_{j=1}^N \delta_D(\vecq - \hat{\vecq}_j(t)) \otimes \vec{e}_j.
\end{align}
Inserting \eqref{eq:collectivePotential} into \eqref{eq:potentialGradient} and going to Fourier space leads us to the interaction operator
\begin{align}\label{eq:interactionOperator}
    \hat{S}_\mathcal{I} = - \int \dd 1 \ \hat{B}(-1) v(1) \hat{\rho}(1),
\end{align}
where we defined the response field operator
\begin{align}\label{eq:responseField}
    \hat{B}(-1)
    \defeq \sum_{j=1}^N \Big(-\ii \veck_1 \cdot \hat{\vec{\chi}}_{p_j} (t_1) \Big)\ee^{\ii \veck_1 \cdot \hat{\vecq}_j(t_1)}
    =\sum_{j=1}^N \hat{b}_j(-\veck_1, t_1) \hat{\rho}_j(-\veck_1, t_1).
\end{align}
%Note that the response field operator contains a one-particle density operator itself and describes the response of the system to a perturbation caused by the presence of an interacting particle \cite{1973PhRvA...8..423M}. 
Note that the response field operator characterizes the response of the system to the presence of an interaction potential.
%The density mode in the response field operator acts as a surrogate for the physical particle that is affected by this perturbation due to two-particle interactions. 
The combinations of density operator and two-particle interaction potential, $v(1) \hat{\rho}(1)$, describes the total interaction potential perturbing the system. Its effect, i.e. the acceleration of particles by the negative potential gradient, is described by the response field.
%The reason for this is that we want to get rid of any trailing factors of $\ii$ in order to make the formulas as transparent as possible. 

%%%%%%%%%%%%%%%%%%%%%%%%%%%%%%%%%%%%%%%%%%%%%%%%%%%%%%%
%%%%%%%%%%%%%%%%%%%%%%%%%%%%%%%%%%%%%%%%%%%%%%%%%%%%%%%
\subsection{Applying operators to the free generating functional}
The free generating functional in \eqref{eq:genFunc2} can be further simplified by inserting the definition of the free action. Solving the $\tensorchi$-integral results in a functional Dirac delta distribution which fixes the phase-space coordinate to follow a classical trajectory with inhomogeneity $\tensorK$. Thus, the free generating functional reads
\begin{align}\label{eq:Simplified_gen_func}
    Z_0[\tensorJ,\tensorK] = \int \dgamma \exp\big( \ii \tensorJ \cdot \tilde{\tensorx}[\tensorK] \big),
\end{align}
with the phase-space trajectory
\begin{equation}\label{eq:particleTrajectories}
	\tilde{\tensorx}[\tensorK](t) = \boldsymbol{\mathcal{G}}(t,0) \, \tensorxin - \int_0^\infty \dd\bar{t} \ \boldsymbol{\mathcal{G}}(t,\bar{t}) \,\tensorK(\bar{t}).
\end{equation}
Here $\boldsymbol{\mathcal{G}}(t,\bar{t}) = \mathcal{G}(t,\bar{t}) \otimes \mathcal{I}_N$ is the Green's function that solves the free equations of motion $\frac{\delta S_0}{\delta \tensorchi} = 0$ for the entire particle ensemble. 
For a single particle the Green's function is a $6 \times 6$ block matrix that reads
\begin{align*}
    \mathcal{G}(t,\bar{t}) = 
    \begin{pmatrix}
    g_{qq}(t, \bar{t})\,\mathcal{I}_3 & g_{qp}(t, \bar{t})\,\mathcal{I}_3
    \\
    g_{pq}(t, \bar{t}) \, \mathcal{I}_3                & g_{pp}(t, \bar{t})\,\mathcal{I}_3
    \end{pmatrix}
    \Theta (t - \bar{t}).
\end{align*}
The functional form of the individual components is determined by the equations of motion. % $\HH_0$,  
The generating functional enables us to generate macroscopic expectation values by application of the corresponding operators. 
Most notably applying density operators to the free generating functional results in a density correlation function $G^{(0)}_{\rho \dotsc \rho}$ evaluated along inertial particle trajectories. 
The effect on the free generating functional is the generation of a shift of the source field \cite{2016NJPh...18d3020B}
\begin{comment}
\begin{align}\label{eq:densityCumulantKFT}
    \big\langle \rho(1) \dotsc \rho(n) \big\rangle = \hat{\rho}(1) \dotsc \hat{\rho}(n) Z_0[\tensorJ,\tensorK] \Big|_{\tensorJ=0} = \sum_{j_1, \dotsc j_n} Z_0[\tensorL, \tensorK],
\end{align}
\end{comment}
\begin{align}\label{eq:densityCumulantKFT}
    G^{(0)}_{\rho \dotsc \rho}(1, \dotsc ,n) 
    %G^{(0)}_{\rho(1) \dotsc \rho(n)}
    = \hat{\rho}(1) \dotsc \hat{\rho}(n) Z_0[\tensorJ,\tensorK]  = \sum_{j_1, \dotsc j_n} Z_0[\tensorJ+\tensorL, \tensorK],
\end{align}
where the shift tensor is given by 
\begin{align}\label{eq:general shift tensor}
    \tensorL(t) = 
    \sum_{s=1}^n 
    \vec{L}_s(t)
    \otimes \vec{e}_{j_s}
    =
    - \sum_{s=1}^n 
    \begin{pmatrix}
    \veck_s \\ 0
    \end{pmatrix}
    \delta_D(t - t_s)
    \otimes \vec{e}_{j_s}.
\end{align}
In the sum over particle indices in \eqref{eq:densityCumulantKFT} each index can represent any particle. This also includes the case where multiple indices take the same value, which is referred to as a shot-noise term. In applications to cosmology, where the number of particles in any volume of interest can be assumed to be extremely large, such terms become sub-dominant and can be neglected \cite{2016NJPh...18d3020B}.
Sums over particle indices are therefore implied as sums over distinct particles.

Perturbations to free correlation functions due to particle interactions can be calculated by applying the interaction operator to the generating functional. 
To this end, the effect of the response field operator has to be investigated in some more detail by evaluating the free correlation function 
%$G^{(0)}_{B\rho \dotsc \rho}$. Applying a single response field operator $\hat{B}(-1')$ onto an $n$-point correlation function we get
\begin{align}\label{eq:responseFieldApplication}
    G^{(0)}_{B\rho \dotsc \rho} (-1'; 1, \dotsc n)
    %G^{(0)}_{B (-1')\rho(1) \dotsc \rho(n)}
    &=
    \hat{B}(-1')\hat{\rho}(1)\dotsc \hat{\rho}(n) Z_0[\tensorJ,\tensorK] \Big|_{\tensorJ=0}\nonumber \\[1.5mm]
    &=
    \sum_{j_0}  \hat{b}_{j_0}(-1') \hat{\rho}_{j_0}(-1') 
    \sum_{j_1, \dotsc j_n} \hat{\rho}_{j_1}(1)  \dotsc \hat{\rho}_{j_n}(n) Z_0[\tensorJ,\tensorK] \Big|_{\tensorJ=0}\nonumber
    \\[1.5mm]
    %&=
    %\sum_{j_0, \dotsc j_n} 
    %\Bigg(-\veck'_{1} \cdot \frac{\delta}{\ii \delta \vecK_{p_{j_0}} (t'_1)} \Bigg) Z_0[\tensorL,\tensorK]\nonumber
    %\\[1.5mm]
    &=
    \sum_{j_0, \dotsc j_n} \Bigg[ \sum_{r=1}^n \int_0^{t_s} \dd \bar{t} \
    \big(-\ii \veck'_1 \cdot \veck_r\big) \delta_D(t'_1 - \bar{t}) \delta_{j_r \, j_0} g_{qp}(t_r,\bar{t})\Bigg] Z_0[\tensorL,\tensorK]\nonumber
    \\[1.5mm]
    %&=
    %\sum_{r=1}^n \big(-\veck'_1 \cdot \veck_r\big) g_{qp}(t_r,t'_1)
    %\ \sum_{j_0, \dotsc j_n} \delta_{j_r \, j_0}Z_0[\tensorL,\tensorK] \nonumber
    %\\
    &=
    \sum_{r=1}^n \ b(r, -1')
    \sum_{j_0, \dotsc j_n} \delta_{j_r \, j_0}Z_0[\tensorL,\tensorK]. %\\[1.5mm]
    %&=
    %\sum_{r=1}^n \ b(r, -1')
    %G_{\rho \dotsc \rho} (1 \dotsc  r-1' \dotsc n).
\end{align}
We have started by expanding the sums over particle indices contained in the individual operators. 
To get to line 3 we applied all one-particle density operators to the generating functional, thus generating a shift tensor $\tensorL$ that is implicitly dependent on the particle indices $j_0, \dotsc j_n$. 
Additionally the response field, containing a functional derivative w.r.t.\ $\vecK_p$, generates the prefactor in the third line. 
Note that, by definition, the function $g_{qp}(t,t')$ vanishes for equal times $t=t'$, such that the sum over $r$ does not include the index 0.
In the last step we rearranged the terms to highlight the effect of the response field and defined the general response field prefactor
\begin{align*}
    b(r,-s') \defeq \big(-\ii\veck'_s \cdot \veck_r\big) g_{qp}(t_r,t'_s).
\end{align*}
Due to the response field, the sum over particle indices is constrained by the Kronecker-delta which sets the particle index of the response field equal to one of the $n$ particle indices of the density fields. 

Generalizing to multiple response fields is straightforward, each of them leading to an additional prefactor and identifying a pair of particle indices.
For a large number of response fields analytical expressions for the resulting terms quickly become unhandy, motivating a diagrammatic approach that was developed and discussed in depth in \cite{2017NJPh...19h3001B}.
The sum in \eqref{eq:responseFieldApplication} then turns into a sum over distinct diagrams weighted with a combinatorial prefactor.

Alternatively, \eqref{eq:responseFieldApplication} can be understood in terms of contractions of response field operators with density operators. Since each particle index is carried by a single density field, setting such an index equal to the index of a response field can be represented as a contraction of the corresponding density field with the response field \cite{2015PhRvE..91f2120V}.
We can then express \eqref{eq:responseFieldApplication} as
\begin{align}\label{eq:responseFieldContraction}
    \hat{B}(-1')\hat{\rho}(1)\dotsc \hat{\rho}(n) Z_0[\tensorJ,\tensorK] \Big|_{\tensorJ=0}
    &=
    \sum_{r=1}^n 
    \wick[offset = 1.2em]
    {\c1{\hat{B}}(-1') \hat{\rho}(1)\dotsc \c1{\hat{\rho}}(r)} 
    \dotsc \hat{\rho}(n) Z_0[\tensorJ,\tensorK]\Big|_{\tensorJ=0}.
    %\\
    %&=
    %\sum_{r=1}^n \ b(r, -1')
    %G_{\rho \dotsc \rho} (1 \dotsc  r-1' \dotsc n),
\end{align}

Applying multiple response fields then leads to a larger number of combinations of contractions or, alternatively, to a larger number of distinct diagrams contributing to the resulting correlation function.
{Furthermore \eqref{eq:responseFieldContraction} can be expressed in terms of free density correlation functions by
\begin{align}\label{eq:free response correlation}
    G^{(0)}_{B\rho \dotsc \rho}(-1'; 1, \dotsc , n) 
    %G^{(0)}_{B (-1')\rho(1) \dotsc \rho(n)}
    = \sum_{r=1}^n \ b(r, -1') 
    G^{(0)}_{\rho \dotsc \rho} (1, \dotsc , r-1', \dotsc ,n),
    %G^{(0)}_{\rho\dotsc \rho}(\wick[offset = 1.0em]
    %{\c1{-1'} ; 1\dotsc \c1{r}}\dotsc n),
\end{align}
where $r-1'$ signifies that the $r$-component of the resulting shift tensor \eqref{eq:general shift tensor} is given by
\begin{align*}
    \vec{L}_r(t) = \begin{pmatrix}
    - \veck_r \\ 0
    \end{pmatrix}
    \delta_D(t - t_r)
    +
    \begin{pmatrix}
    \veck'_1 \\ 0
    \end{pmatrix}
    \delta_D(t - t'_1).
\end{align*}
Considering the definition \eqref{eq:interactionOperator} of the interaction operator, \eqref{eq:free response correlation} implies that an interacting $n$-point density function is dependent on all free density correlation functions of higher order, a statement equivalent to the BBGKY hierarchy. This can be seen by expanding the exponential of the interaction operator into a series, applying it to the free generating functional and expressing the result in terms of free correlation functions \cite{2015PhRvE..91f2120V}.}
%\subsection{Comparison to other approaches to cosmic structure formation}

%%%%%%%%%%%%%%%%%%%%%%%%%%%%%%%%%%%%%%%%%%%%%%%%%%%%%%%
%%%%%%%%%%%%%%%%%%%%%%%%%%%%%%%%%%%%%%%%%%%%%%%%%%%%%%%
\subsection{Evaluation of the free generating functional}
Once the density and response field operators have been applied, the key object that needs to be evaluated to calculate statistical quantities in KFT is the free generating functional.
For any order in the perturbative expansion of the interaction operator, the final expression is given in terms of space-time integrals over the free generating functional $Z_0[\tensorL,0]$, where the shift tensor $\tensorL$ is determined by the operators that were applied to $Z_0$.
Independently of the exact form of the shift tensor, the free generating functional evaluated at $\tensorK = 0$ reads
\begin{align*}
    Z_0[\tensorL,0] = \int \dgamma \ \ee^{\ii \tensorL \cdot \bar{\tensorx}},
\end{align*}
where $\bar{\tensorx}(t) = \tilde{\tensorx}[\tensorK] \big|_{\tensorK = 0}$ is the inertial part of the particle trajectory \eqref{eq:particleTrajectories}. 
Since we average over initial phase-space coordinates, it is useful to collect the respective prefactors of $\tensorqin$ and $\tensorpin$ and rewrite the phase factor such that
\begin{align}\label{eq:Z0Phase}
    \ii \tensorL \cdot \bar{\tensorx}
    %&= 
    %\ii \sum_{s=1}^n \vec{L}_{q_s} \cdot \big( {\vecqin}_{s} + g_{qp}(t_s, 0)\vecpin_s \big)
    %\\
    %&\eqdef \ii \sum_{s=1}^n \vec{L}_{q_s} \cdot {\vecqin}_{s} +
    %\ii \sum_{s=1}^n \vec{L}_{p_s} \cdot {\vecpin}_{s}
    %\\
    &\eqdef 
    \ii \tensorLq \cdot \tensorqin + \ii \tensorLp \cdot \tensorpin.
\end{align}
We call $\tensorLq$ and $\tensorLp$ position and momentum shifts, but note that they are not equivalent to the respective components of the shift tensor $\tensorL$.
In the case where $n$ density and no response field operators have been applied to the free generating functional, the position and momentum shifts are given by
\begin{align}\label{eq:qpShiftTensors}
\tensorLq = - \sum_{s=1}^n \veck_{s} \otimes \vec{e}_{j_s}
\quad ,   \qquad
\tensorLp = - \sum_{s=1}^n g_{qp}(t_s,0) \veck_s \otimes \vec{e}_{j_s}.
\end{align}
As was shown in \eqref{eq:responseFieldApplication}, applying response field operators identifies pairs of particle indices, thereby changing the sums in \eqref{eq:qpShiftTensors}. 
This has been discussed in more detail in \cite{2016NJPh...18d3020B, 2017NJPh...19h3001B}.
Employing \eqref{eq:Z0Phase}, the free generating functional takes the simple form
\begin{align}\label{eq:Z0 in terms of Lq and Lp}
    Z_0[\tensorL,0] = \int \dgamma \ \ee^{\ii \tensorLq \cdot {\tensorqin} + \ii \tensorLp \cdot \tensorpin}.
\end{align}

Even though $Z_0$ contains no interactions, it should be stressed that its evaluation is far from trivial. This is due to the fact that KFT encodes the complete information on the initial correlations of the system at hand in the free generating functional. 
In contrast, all information on particle interactions is contained in the interaction operator \eqref{eq:interactionOperator}. 
In comparison to more established analytical methods for cosmic structure formation, such as SPT and LPT, KFT thus offers a particularly transparent scheme to analyze and interpret the respective effects of gravitational interactions and initial correlations separately.

As described previously, the information on the initial conditions is contained in the integration measure
$\dgamma = \dd \tensorxin P(\tensorxin)$,
where $P(\tensorxin)$ is the probability distribution underlying the initial phase-space coordinates of the classical point particles of the system.
It is derived from the known initial state of the density and velocity fields via Poisson sampling
%, which is discussed in great detail in 
\cite{2016NJPh...18d3020B, 2021JCAP_Kozlikin}.
The resulting probability distribution is written in terms of a covariance matrix containing density-density, density-momentum as well as momentum-momentum correlations. 

In our application to cosmic structure formation we introduce Zel'dovich trajectories as inertial reference trajectories such that the particle propagator becomes large for late cosmic times, allowing us to neglect all but initial momentum-momentum correlations \cite{2016NJPh...18d3020B}.
Note that this assumption is in no way necessary for KFT, but leads to considerable simplifications in the evaluation of the free generating functional \cite{2018JSMTE..04.3214F}.
With the assumption of vanishing initial density-density and density-momentum correlations the initial phase-space probability distribution can be shown \cite{2017NJPh...19h3001B} to read
\begin{align*}
    P(\tensorxin) = \frac{V^{-N}}{\sqrt{\det(2\pi \mathbf{C}_{pp})}} \exp\Big(-\tfrac{1}{2} \boldsymbol{p}^{(\ii)\,\mathrm{T}} \cdot \mathbf{C}_{pp}^{-1} \cdot \tensorpin \Big).
\end{align*}
At first glance this looks like the initial phase-space coordinates simply follow a Gaussian probability distribution.
One should keep in mind, however, that the momentum covariance matrix 
\begin{align}\label{eq:Cpp}
    \mathbf{C}_{pp} = \sum_{i,j} C_{pp} \big(q^{\mathrm{(i)}}_{ij}\big) 
    \otimes (\vec{e}_i \otimes \vec{e}_j)
\end{align}
is a non-trivial function of the distance $q^{\mathrm{(i)}}_{ij} = |\vecq^{\, \mathrm{(i)}}_{ij}|=|\vecqin_i - \vecqin_j|$ between pairs of particles and relates to the initial density fluctuation power spectrum through
\begin{align}\label{eq:Cpp in terms of power spectrum}
    C_{pp} \big(q^{\mathrm{(i)}}_{ij}\big) = \int_k \frac{\veck \otimes \veck}{k^4} P_\delta^{ \mathrm{(i)}}(k) \ee^{\ii \veck \cdot \vecq^{\,\mathrm{(i)}}_{ij}}.
\end{align}
While the dependence of the initial probability distribution on particle momenta is indeed given by a simple Gaussian function, initial positions enter in a way that renders the evaluation of \eqref{eq:Z0 in terms of Lq and Lp} rather complicated.
Making use of the fact that the momentum covariance matrix depends only on the distance between pairs of particles, a factorization scheme was introduced in \cite{2017NJPh...19h3001B}.
The resulting expression for the free generating functional with an $n$-point shift tensor $\tensorL$ reads
%%%%%%%%%%%%%%%%%%%%%%%%%%%%%%%%%%%%%%%%%%%%%%%
\begin{comment}
\begin{align}\label{eq:factorizedZ}
    Z_0[\tensorL,0] = 
    (2\pi)^3 V^{-n} 
    \delta_D\Big(\sum_{j=1}^n \vecL_{q_j}\Big)
    e^{-Q_D} \prod_{2\leq b < a}^n 
    \int_{k'_{ab}} \prod_{1\leq i < j}^n 
    \left(\curlyP_{ij}(\veck_{ij}) + \Delta_{ij} (\veck_{ij}) \right),
\end{align}
where the factors of the generating functional $\mathcal{P}_{ij}$ and $\Delta_{ij}$ are defined by
\begin{align}\label{eq:curlyP}
    \curlyP_{ij}(\veck_{ij})= \int_{q}
    \left(
    e^{-\vecL^{\,\mathrm{T}}_{p_i} C_{p p}(q) \vecL_{p_j}}
    -1\right)
    e^{\ii\veck_{ij} \cdot \vecq}
    \quad , \qquad
    \Delta_{ij}= (2\pi)^3 \delta_D(\veck_{ij}).
\end{align}
\end{comment}
%%%%%%%%%%%%%%%%%%%%%%%%%%%%%%%%%%%%%%%%%%%%%%%
\begin{align}\label{eq:factorizedZ}
    Z_0[\tensorL,0] = 
    (2\pi)^3 V^{-n} 
    \delta_D\Big(\sum_{j=1}^n \vecL_{q_j}\Big)
    e^{-Q_D} \prod_{2\leq b < a}^n 
    \int_{k'_{ab}} \prod_{1\leq i < j}^n 
    \left(\curlyP_{ij}(\veck_{ij}) +(2\pi)^3 \delta_D(\veck_{ij}) \right),
\end{align}
where the factors of the generating functional $\mathcal{P}_{ij}$ are defined by
\begin{align}\label{eq:curlyP}
    \curlyP_{ij}(\veck_{ij})= \int_{q}
    \left(
    e^{-\vecL^{\,\mathrm{T}}_{p_i} C_{p p}(q) \vecL_{p_j}}
    -1\right)
    e^{\ii\veck_{ij} \cdot \vecq}.
\end{align}
Since the covariance matrix depends only on the norm of $\vecq$, we have $\curlyP_{ij}(-\veck_{ij}) = \curlyP_{ij}(\veck_{ij})$.
To bring the expression into a form that can be factorized, auxiliary wave-vectors $\veck_{ij}$ were introduced such that
\begin{align*}
    \veck_{ij}=
    \begin{cases}
    \vecL_{q_i}- \sum_{b=2}^{i-1} \veck'_{ib} + \sum_{a=i+1}^{n} \veck'_{ai} \qquad &j=1
    \\
    \veck'_{ij}  &j\neq 1.
    \end{cases}
\end{align*}
The exponential factor in front of the integral is a Gaussian {damping factor} attributed to the variance of the initial momentum distribution
\begin{align}\label{eq:damping}
    Q_D = \frac{\sigma_p^2}{2}\sum_{j=1}^n \vecL_{p_j}^{\,2},
\end{align}
where $\sigma_p^2 = \frac{1}{3}\int_k \frac{P_\delta(k)}{k^2}$ is the momentum dispersion.
It was shown in \cite{2006PhRvD..73f3519C} that such a damping (or smearing) can be obtained in SPT from the resummation of special sub-classes of diagrams.
%It is well known that such a damping (or \textit{smearing}) can not be derived in purely perturbative frameworks such as SPT \cite{2008PhRvD..77f3530M}, since any truncation in initial correlations can only lead to a power series of finite order in $\sigma_v^2$.
%In fact, in order to just arrive at this exponential damping in SPT an infinite class of sub-diagrams needs to be summed up, leading to renormalized perturbation theory (RPT) \cite{2006PhRvD..73f3519C, 2006PhRvD..73f3520C}.
In free KFT the damping factor arises in a natural way due to the inclusion of the full hierarchy of momentum correlations in the initial conditions\footnote{It was shown recently \cite{2021-2110.07427_Konrad, 2022-2202.08059_Konrad} that, in the KFT formalism, a free power spectrum attains an asymptotic $k^{-3}$-behaviour for large wave-numbers $k$ due to an exact cancelling of the leading order asymptotic expansion of $\curlyP$ and the exponential damping factor.}.

It is easy to check, and we will see below, that \eqref{eq:factorizedZ} reproduces the well-known Zel'dovich power spectrum \cite{2006PhRvD..73f3519C,2008PhRvD..77f3530M} in the special case where we apply two density operators ($n=2$) at equal time to $Z_0$ and make an appropriate choice for the inertial particle trajectories (see \autoref{sec:particle trajectories}).
%This is the zero-th order result of a KFT power spectrum. 
Applying interactions by Taylor expanding the exponential interaction operator in \eqref{eq:genFunc2} adds correction terms that contain an increasingly large number of shift vectors, resulting in convolutions of $\curlyP$-factors \cite{2017NJPh...19h3001B}. 
We will see this idea explicitly at work in \autoref{sec:perturbation theory}.

\section{Cosmological particle trajectories}
\label{sec:particle trajectories}
In this section we introduce trajectories for classical, non-relativistic dark matter particles in an expanding universe.
{
We should note that efforts to extend our formalism to allow for the description of relativistic particles have, so far, been in vain. 
While one can, in principle, incorporate different particle species such as massive neutrinos after they have become non-relativistic, a description based on the trajectories of such particles in the relativistic regime is so far out of reach.}

For the description of non-relativistic dark matter particles, it is convenient for to replace the cosmic time $t$ with the time coordinate 
\begin{align} \label{eq:eta definition}
    \eta(t) \defeq \ln \frac{D_+(t)}{D_+(t_\ii)},
\end{align}
where $D_+(t)$ is the linear growth factor and $t_\ii$ is the starting time of our calculations, for which we choose the time of the CMB decoupling. Note that with this definition $\eta(t_\ii)=0$.

\subsection{Newtonian and Zel'dovich trajectories}
In \autoref{apx:trajectories} we show that, using $\eta$ as time coordinate, particles in an expanding space time follow phase-space trajectories of the form
\begin{align}\label{eq:particle trajectories Physical}
    \Vec{x}(\eta) = 
    \mathcal{G} (\eta,0)\vec{x}^{\,(\ii)}
    -\int_0^\infty \dd \bar{\eta} \ 
    \mathcal{G} (\eta,\bar{\eta})\vec{\mathcal{F}}(\bar{\eta}),  
\end{align}
with the Green's function
\begin{align}\label{eq:Greens functions}
     \mathcal{G}(\eta,\bar{\eta}) = 
    %\begin{pmatrix}
    %\mathcal{I}_3 g_{qq}(\eta,\bar{\eta}) 
    %& \mathcal{I}_3 g_{qp}(\eta,\bar{\eta})
    %\\
    %\mathcal{I}_3 g_{pq}(\eta,\bar{\eta})                  
    %& \mathcal{I}_3 g_{pp}(\eta,\bar{\eta})
    %\end{pmatrix} 
    %\Theta (\eta - \bar{\eta})
    %=
     \begin{pmatrix}
    \mathcal{I}_3 & 2\left(1-\ee^{-\frac{1}{2}(\eta-\bar{\eta})}\right)\mathcal{I}_3
    \\
    0_3                  & \ee^{-\frac{1}{2}(\eta-\bar{\eta})}\mathcal{I}_3
    \end{pmatrix}
    \Theta (\eta - \bar{\eta}).
\end{align}
Since we chose the phase-space coordinates such that $\vec{p}(\eta) = \Dot{\vec{q}}(\eta)$, the components of the Green's function are related by $g_{pp}(\eta,\bar{\eta}) =  \dot{g}_{qp}(\eta,\bar{\eta})$ and $g_{pq}(\eta,\bar{\eta}) = \dot{g}_{qq}(\eta,\bar{\eta})$, where we defined the derivative w.r.t.\ $\eta$ by a dot. 
Since $g_{qq}(\eta,\bar{\eta})=1$, the sole important component of the Green's function is 
$g_{qp}(\eta,\bar{\eta})$, which we call the \textit{propagator}.
The inhomogeneity $\vec{\mathcal{F}} = \big(0, \; {\vnabla}\phi \big)^\mathrm{T}$ arises due to particle interactions, which are collectively described by the potential $\phi$, satisfying the Poisson equation
\begin{align}\label{eq:Poisson equation}
    \vnabla^{\,2} \phi (\vecq, \eta) = \frac{3}{2} \delta (\vecq, \eta),
\end{align}
where $\delta(\vecq, \eta)$ is the density contrast. 
The resulting interactions between individual particles are mediated by a Newtonian potential.
Since we are using comoving coordinates, free Newtonian motion is bounded, which is reflected by the fact that $\lim_{\eta\rightarrow \infty} g_{qp}(\eta,0) = 2 < \infty$. 
This is problematic for our approach to KFT, since the free-streaming growth of the power spectrum is directly related to $g_{qp}^2(\eta,0)$ \cite{2016NJPh...18d3020B, 2017NJPh...19h3001B}. 
Using Newtonian trajectories as defined above would require us to include an infinite subset of interaction contributions in order to recover the linear growth of the dark matter power spectrum \cite{Lilow_2019}.

To describe the growth of large-scale structures already with non-interacting KFT, we impose inertial\footnote{In this context `inertial' denotes the part of the particle trajectories that can be described as a time-dependent linear transformation on the initial phase-space coordinates. We refrain from referring to it as `free motion' since it may already incorporate gravitational effects, e.g.\ those that lead to linear structure growth in the case of Zel'dovich trajectories.} 
Zel'dovich trajectories
\begin{align*}
    \vec{q}(\eta) = \vecqin + \big(\ee^\eta -1\big) \ \vecpin .
    %\qquad \mathrm{with} \quad g_{qp}^{(\mathcal{Z})} (\eta) = \ee^\eta -1.
\end{align*}
Intuitively this definition shifts part of the particle interactions---namely the part that is responsible for linear growth---into the inertial motion of particles.
In \autoref{apx:trajectories} we show that the resulting single-particle phase-space trajectories read
\begin{equation}\label{eq:particleTrajectories Zeldovich}
	\vecx(\eta) = {\mathcal{G}}_\mathcal{Z}(\eta,0) \, \vecxin - \int_0^\infty \dd\bar{\eta} \ {\mathcal{G}}(\eta,\bar{\eta}) \,\vec{\mathcal{F}}_\mathcal{Z}(\bar{\eta}),
\end{equation}
where the inhomogeneity $\vec{\mathcal{F}}_\mathcal{Z} = \big(0, \; {\vnabla}V \big)^\mathrm{T}$ is defined in terms of a potential fulfilling the modified Poisson equation
\begin{align} \label{eq:mod Poisson_eq}
    \vnabla^{\,2} V (\vecq, \eta) = \frac{3}{2} \left(\delta (\vecq, \eta)- \ee^\eta \delta^{(\ii)}(\vec{q}) \right).
\end{align}
Deviations from inertial trajectories are thus sourced by density fluctuations relative to the linearly evolved density field. 
With this modification the contribution of particle interactions that leads to linear growth, and which was shifted into the inertial motion of particles when imposing Zel'dovich trajectories, is now subtracted from the interaction potential.
The Green's function for the inertial part of the trajectories is defined as 
\begin{align}\label{eq:Greens function Zeldovich}
     \mathcal{G}_\mathcal{Z}(\eta,0) = 
    \begin{pmatrix}
    \mathcal{I}_3 & g_{qp}^{(\mathcal{Z})} (\eta,0)\,\mathcal{I}_3
    \\
    0_3           & \dot{g}_{qp}^{(\mathcal{Z})}  (\eta,0)\, \mathcal{I}_3
    \end{pmatrix} 
    \Theta(\eta - \bar{\eta}),
    \qquad \mathrm{with} \quad g_{qp}^{(\mathcal{Z})} (\eta,0) =\big( \ee^{\eta} -1\big).
\end{align}
We want to emphasize that the trajectories \eqref{eq:particleTrajectories Zeldovich} and \eqref{eq:particle trajectories Physical} are completely equivalent, since they can be transformed into each other by following the procedure in \autoref{apx:trajectories}. 
Nevertheless, when treating interactions perturbatively in the KFT framework, expression \eqref{eq:particleTrajectories Zeldovich} is preferable, since large-scale structure growth is contained in the inertial motion of particles already.
This is not intrinsically required by KFT, however; in fact, the growth of cosmological large-scale structures can be obtained directly from Newtonian particle dynamics by employing the resummation scheme presented in \cite{Lilow_2019}.

\subsection{Incorporating Zel'dovich trajectories into KFT}
To include the particle trajectories \eqref{eq:particleTrajectories Zeldovich} into the KFT formalism, let us first take a look at the interaction operator from a different angle by rewriting \eqref{eq:interaction operator 1} as
\begin{equation}
    \hat{S}_\mathcal{I}
    = \boldsymbol{\mathcal{F}}(\hat{\tensorq}) \cdot 
    \frac{\delta}{\ii \, \delta \tensorK},
\end{equation}
where we introduced $\boldsymbol{\mathcal{F}}(\hat{\tensorq}) \defeq \big(0, \; \boldsymbol{\nabla}V(\hat{\tensorq}) \big)^\mathrm{T}$.
In this form the exponential of the interaction operator, similarly to a density operator, generates the shift $\tensorK \rightarrow \tensorK+ \boldsymbol{\mathcal{F}}$ in the free generating functional. 
Setting the source field $\tensorK$ to 0 afterwards, we recover the physically exact particle trajectories in the generating functional, illustrating that the application of the exponential interaction operator to the free generating functional leads to the full physical particle trajectories, as expected.

Conversely, this implies that combinations of interaction operator $\hat{S}_\mathcal{I}$ and particle trajectory $\tilde{\tensorx}[\tensorK]$ resulting in the same trajectory $\tilde{\tensorx}[\boldsymbol{\mathcal{F}}]$ are equivalent.
The equivalence of pure Newtonian phase-space trajectories \eqref{eq:particle trajectories Physical} and trajectories with inertial Zel'dovich motion \eqref{eq:particleTrajectories Zeldovich} then implies that the combination
\begin{align}
    \tilde{\tensorx}[\tensorK](\eta) &= \boldsymbol{\mathcal{G}}_\mathcal{Z}(\eta,0) \, \tensorxin - \int_0^\infty \dd\bar{\eta} \ \boldsymbol{\mathcal{G}}(\eta,\bar{\eta}) \,\tensorK(\bar{\eta}),
    \\
    \hat{S}_\mathcal{I} &= \boldsymbol{\mathcal{F}}_\mathcal{Z}(\hat{\tensorq}) \cdot 
    \frac{\delta}{\ii \, \delta \tensorK}, \label{eq:interaction operator Zel'dovich}
\end{align}
also leads to the exact physical trajectories \eqref{eq:particle trajectories Physical}.
In this case, however, inertial particle motion is given by the Zel'dovich propagator, whereas particle interactions appear together with the propagator of the Hamiltonian equations of motion.
{Of course, shifting around contributions from particle interaction to inertial motion is not restricted to inertial Zel'dovich motion. 
In principle, one has unlimited choice in defining inertial phase-space trajectories, as long as the corresponding particle interactions are chosen consistently.
The final aim of this construction is to find inertial particle trajectories that minimize the impact of particle interactions on the non-linear evolution of cosmic structures.
In \cite{PhysRevD.97.023508_McDonald_Vlah}, where a formalism similar to KFT is introduced, the authors enforce Zel'dovich trajectories in a similar fashion, but additionally include a scale-dependent damping factor into the propagator.
Alternatively, implementing inertial particle trajectories that mimic the adhesion approximation \cite{10.1093/mnras/236.2.385_adhesion} would in principle be possible, although one has to keep in mind that KFT works with the phase-space trajectories of individual particles whereas the adhesion approximation is usually employed in the context of fluid dynamics.
In any case, constructions of this sort are easily implemented in KFT but are yet to be explored in more detail. 
}

Returning to the collective potential that appears in the interaction operator \eqref{eq:interaction operator Zel'dovich} we note that it is defined by the modified Poisson equation \eqref{eq:mod Poisson_eq}.
However, since KFT describes a system on the basis of individual particles and their interactions, an expression for the two-particle interaction potential is needed.
It was shown in \cite{2020arXiv_MeanField} that \eqref{eq:mod Poisson_eq} implies a two-particle interaction potential $v(\veck, \eta)$ that can be approximated by
\begin{align}\label{eq:modified two-particle potential}
    v(\veck, \eta) = -\frac{3}{2\bar{\rho}} \frac{1}{k^2}
    f_v(k,\eta), 
    \quad 
    \mathrm{with}
    \qquad
    f_v(k,\eta) = 
    1-\sqrt{\frac{P_\delta^{(\mathrm{lin})}(k,\eta)}{P_\delta(k,\eta)}},
\end{align}
where $P_\delta^{(\mathrm{lin})}(k,\eta)$ and $P_\delta(k,\eta)$ are the linear and the non-linear dark matter power spectra respectively.
The function $f_v(k)$ can be determined approximately from first order KPT, which is discussed in \autoref{apx:Yukawa cutoff}.
It is shown there that we can approximate the two-particle interaction potential for particles moving along inertial Zel'dovich trajectories by a Yukawa potential
\begin{align*}
    v(\veck, \eta) = -\frac{3}{2\bar{\rho}} \frac{1}{k^2 + k_0^2(\eta)},
\end{align*}
with the time-dependent Yukawa cutoff $k_0(\eta)$ given by \eqref{eq:Yukawa cutoff apx}.

\section{Perturbative expansion for a power spectrum}
\label{sec:perturbation theory}
The full non-linear density fluctuation power spectrum $P_\delta(k_1,\eta_1)$ is related to the equal-time two-point function $G_{\rho\rho} (1, \,2)\big|_{\eta_1= \eta_2}$ by
\begin{align}\label{eq:non-linear PS}
    P_\delta(k_1,\eta_1) = \frac{1}{\Bar{\rho}^{\,2}} \int_{k_2} G_{\rho\rho} (1, \,2)\Big|_{\eta_1= \eta_2} - (2\pi)^3 \delta_D(\veck_1),
\end{align}
where the term $\hat{1} = (2\pi)^3 \delta_D(\veck_1)$ is subtracted, to get from a density to a density fluctuation correlation function. 
The full two-point function $G_{\rho\rho} (1, \,2)$ can be obtained from KFT by applying two density operators to the fully interacting generating functional
\begin{align}\label{eq:two-point function}
    G_{\rho\rho} (1, \,2)\Big|_{\eta_1= \eta_2} 
    = \hat{\rho}(1) \hat{\rho}(2)\Big|_{\eta_1= \eta_2} Z[\tensorJ,\tensorK]\Big|_0 
    = \hat{\rho}(\veck_1, \eta_1) \hat{\rho}(\veck_2, \eta_1) Z[\tensorJ,\tensorK]\Big|_0 .
\end{align}
%In the following the evaluation at equal times $\eta_1 = \eta_2$ will always be implied.
Evaluating the full generating functional \eqref{eq:genFunc2} in \eqref{eq:two-point function} is unfeasible, since the interaction operator cannot be evaluated exactly. 
To make further progress we expand the exponential of the interaction operator into a power series, leading to a sum of correction terms to the free generating functional
\begin{align}
    Z[\tensorJ,\tensorK] = \exp(\ii\hat{S}_\mathcal{I}) Z_0[\tensorJ, \tensorK] = \Big(1 +\ii\hat{S}_\mathcal{I} + \tfrac{1}{2}(\ii \hat{S}_\mathcal{I})^2 + \mathcal{O}(\hat{S}_\mathcal{I}^3) \Big) Z_0[\tensorJ, \tensorK],
\end{align}
and truncate the sum at second order. 
Since the interaction operator introduces deviations from inertial particle trajectories due to the presence of other particles, its effect is small on large scales, where Zel'dovich trajectories model the growth of structures extremely well. 
As we go towards smaller scales, interactions between particles become more dominant, and higher orders of the expansion in the interaction operator are needed to describe the growth of structures accurately.
{ We should note that, at present, we do not make any claim about the convergence of the series expansion of the interaction operator in KFT. 
As we will see, the results suggest that perturbative corrections to the power spectrum do decrease in magnitude as we proceed towards higher orders.
Nonetheless, it is well possible that perturbation theory leads to an asymptotic series that does not converge to the true non-linear result, but approximates it well enough for finite orders.}

{We also want to stress that, in contrast to other perturbative treatments of structures formation such as SPT or the traditional approach to LPT, KFT does not rely on an expansion in the density contrast $\delta(k)$.}
The two-point function at second-order in the interaction operator is given by
\begin{align}\label{eq:two-point function second order}
    G^{(\leq 2)}_{\rho\rho} (1, \,2) 
    = \hat{\rho}(1) \hat{\rho}(2) Z_0[\tensorJ,\tensorK]\Big|_0 
    + \ii\hat{S}_\mathcal{I} \hat{\rho}(1) \hat{\rho}(2) Z_0[\tensorJ,\tensorK]\Big|_0 
    + \tfrac{1}{2} (\ii \hat{S}_\mathcal{I})^2 \hat{\rho}(1) \hat{\rho}(2)Z_0[\tensorJ,\tensorK]\Big|_0,
\end{align}
where evaluation of the two density operators at equal times $\eta_1 = \eta_2$ is implied.

%%%%%%%%%%%%%%%%%%%%%%%%%%%%%%%%%%%%%%%%%%%%%%
%%%%%%%%%%%%%%%%%%%%%%%%%%%%%%%%%%%%%%%%%%%%%%
%\subsection{Evaluation of perturbative terms}
%%%%%%%%%%%%%%%%%%%%%%%%%%%%%%%%%%%%%%%%%%%%%%
\subsection{Non-interacting two-point function}
Expression \eqref{eq:two-point function second order} is to be evaluated term by term, starting with the lowest order. Evaluating the non-interacting term using \eqref{eq:factorizedZ} results in 
\begin{align*}
    \hat{\rho}(1) \hat{\rho}(2) Z_0[\tensorJ,\tensorK]\Big|_0 
    =
    (2\pi)^3 \Bar{\rho}^2 
    \delta_D\big(\veck_1 + \veck_2\big)
    \ee^{-Q_D(\tensorL_p)} 
    \big[
    \curlyP_{21} (\veck_1) 
    + (2\pi)^3 \delta_D(\veck_2)
    \big],
\end{align*}
with momentum and position shift vectors given by
\begin{align*}
    \vecL_{q_1} &= - \veck_1, \qquad \qquad\vecL_{p_1} = - g_{qp}^{(\mathcal{Z})}(\eta_1,0) \veck_1,
    \\
    \vecL_{q_2} &= - \veck_2, \qquad \qquad \vecL_{p_2} = - g_{qp}^{(\mathcal{Z})}(\eta_1,0) \veck_2.
\end{align*}
{With these definitions for the shift vectors} the factor
\begin{align*}
    \curlyP_{21} (\veck_1) = \int_{q}
    \left(
    \ee^{-\vecL^{\,\mathrm{T}}_{p_1} C_{p p}(q) \vecL_{p_2}}
    -1\right)
    \ee^{\ii\veck_1 \cdot \vecq},
\end{align*}
together with the damping factor, reproduces the well-known Zel'dovich power spectrum \cite{2006PhRvD..73f3519C, 2006PhRvD..73f3520C}.
\begin{comment}
, such that \eqref{eq:non-linear PS} becomes
\begin{align*}
    P^{(0)}_\delta(k_1,\eta_1) =  \ee^{-Q_D} \int_{q}
    \left(
    \ee^{-\vecL_{p_1} C_{p p}(q) \vecL_{p_2}}
    -1\right)
    \ee^{\ii\veck_1 \cdot \vecq},
\end{align*}
\end{comment}

%%%%%%%%%%%%%%%%%%%%%%%%%%%%%%%%%%%%%%%%%%%%%%
\subsection{First-order corrections}
\begin{figure}
    \centering
    \includegraphics[scale = 0.9, clip=true, trim=5cm 20cm 9cm 0]{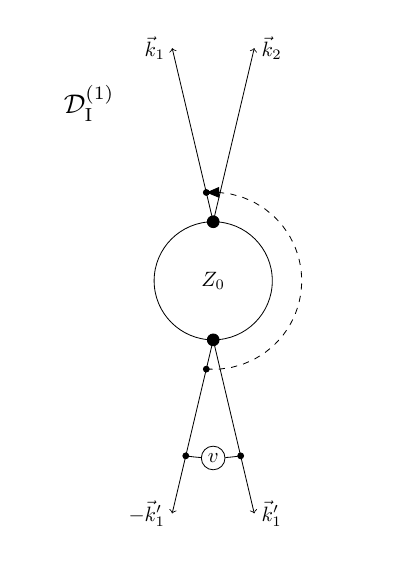}
    \caption{Diagram corresponding to the first order contribution \eqref{eq:first order contraction}. According to the rules of \cite{2017NJPh...19h3001B}, we attribute to this diagram a multiplicity of 2, since connecting to one or the other external mode $\veck_1$ or $\veck_2$ yields the same result for an equal-time two-point correlator in an isotropic random field. }
    \label{fig:first order diagram}
\end{figure}

To go beyond the Zel'dovich power spectrum we need to evaluate the second term in \eqref{eq:two-point function second order}, containing the interaction operator at first order. This expression has been evaluated in \cite{2016NJPh...18d3020B}, albeit with a less rigorous treatment of the particle interaction potential and using a modified version of Zel'dovich trajectories. 
We will therefore give a detailed summary of the derivation.
Since the interaction operator \eqref{eq:interactionOperator} contains a single response field, only a single distinct configuration exists, resulting in
\begin{align}\label{eq:first order contraction}
    %\big\langle \rho(1) \rho(2) \big\rangle^{(1)} &= 
    \ii\hat{S}_\mathcal{I} \ \hat{\rho}(1) \hat{\rho}(2) Z_0[\tensorJ, \tensorK] \Big|_{0}
    &
    =
    %\\
    %&= 
    -\ii \int \dd 1' \ v(1') \hat{B}(-1') \hat{\rho}(1') \ \hat{\rho}(1) \hat{\rho}(2) Z_0[\tensorJ, \tensorK] \Big|_{0} \nonumber
    \\
    &=
    -2\ii \int \dd 1' \ v(1') \wick[offset = 1.2em]{\c1{\hat{B}}(-1') \hat{\rho}(1') \c1{\hat{\rho}}(1) \hat{\rho}(2)} Z_0[\tensorJ, \tensorK] \Big|_{0} 
    \nonumber
    \\
    %&=
    %2 \int \dd 1' \ v(1') b(1,-1') \ \hat{\rho}(1') \hat{\rho}(1, -1') \hat{\rho}(2) Z_0[\tensorJ, \tensorK] \Big|_{\tensorJ = \tensorK = 0}
    %\\
    %&=
    %2 \int \dd 1' \ v(1')  b(1,-1') \ N^3 Z_0[\tensorL, 0]
    &=
    - 2\ii \int \dd 1' \ v(1')  b(1,-1') 
    G^{(0)}_{\rho\rho\rho}(1-1', 2, 1')
    %G^{(0)}_{\rho\rho\rho}(\wick[offset = 1.0em]{\c1{-1'};\c1{1}}, 2, 1')
    . 
\end{align}
This is true in the case where both external density operators are evaluated at equal times (i.e.\ in the case of an equal-time power spectrum), since contraction of the response field with either density mode is equivalent, giving rise to the factor $2$ in the second line.
Note again that there is no contribution from the contraction of the response field with the density $\hat{\rho}(1')$ due to the vanishing propagator $g_{qp}(t_1',t_1')$ in $b(1',-1')$.
In the diagrammatic approach of \cite{2017NJPh...19h3001B} expression \eqref{eq:first order contraction} is represented by \autoref{fig:first order diagram}.
%%%%%%%%%%%%%%%%%%%%%%%%%%%%%%%%%%%%%%%%%
%%%%%%%%%%%%%%%%%%%%%%%%%%%%%%%%%%%%%%%%%
The application of the response field operator influences the shift tensor through the identification of particle indices as seen in \eqref{eq:responseFieldApplication}, resulting in the shift vectors
\begin{align*}
    \vecL_{q_1} &= - \veck_1 + \veck'_1, \qquad \qquad \vecL_{p_1} = - g_{qp}^{(\mathcal{Z})}(\eta_1,0) \veck_1 + g_{qp}^{(\mathcal{Z})}(\eta'_1,0) \veck'_1,
    \\
    \vecL_{q_2} &= - \veck_2, \qquad \qquad \qquad \, \vecL_{p_2} = - g_{qp}^{(\mathcal{Z})}(\eta_1,0) \veck_2,
    \\
    \vecL_{q_3} &= - \veck'_1 , \qquad \qquad \qquad \,  \vecL_{p_3} = - g_{qp}^{(\mathcal{Z})}(\eta'_1,0) \veck'_1.
\end{align*}
This can be directly read off the free three-point function in \eqref{eq:first order contraction}. 
Factorizing the resulting free generating functional and carrying out the integral over the auxiliary wave vector $\veck'_{32}$ where possible yields
\begin{multline}\label{eq:generating functional first order}
   G^{(0)}_{\rho\rho\rho}(1-1', 2, 1')=
    (2\pi)^3 \Bar{\rho}^{3} 
    \delta_D\left(\veck_1+ \veck_2 \right)
    \ee^{-{Q}_D(\tensorL_p)} 
    \Big[
   (2\pi)^3 \delta_D(\veck_1'-\veck_1)\, \curlyP_{32}(\veck_1)
   \\[2mm]
   +
   \curlyP_{21}(\veck_1)\,\curlyP_{31}(\veck_1')
   +
   \curlyP_{21}(\veck_1'-\veck_1)\,\curlyP_{32}(\veck'_1)
   +
   \curlyP_{31}(\veck_1'-\veck_1)\,\curlyP_{32}(\veck_1)
   \\[2mm]
   +
   \int_{k_{32}'}
   \curlyP_{21}(\veck_1+\veck_{32}')\,
   \curlyP_{31}(\veck_1'+\veck_{32}')\,
   \curlyP_{32}(\veck_{32}') \Big].
\end{multline}
At this point we want to emphasise again the conceptual difference between KPT and theories that rely on an expansion in terms of the density contrast. 
Expression \eqref{eq:generating functional first order} inserted in \eqref{eq:first order contraction} describes the full contribution to the first-order correction to the Zel'dovich power spectrum in KFT. 
First order in KFT means that we take into account deviations from inertial particle trajectories at first order, i.e.\ particle trajectories are perturbed due to the presence of other particles moving along inertial trajectories. 
As we go to higher orders, the trajectories of perturbing particles will deviate from inertial trajectories themselves.
On the other hand, since each factor in \eqref{eq:generating functional first order} is of infinite order in the initial correlations, we are not considering an expansion in terms of the density contrast.
This illustrates that KFT decouples the effect of initial correlations from interactions between components of the system.
%Unfortunately a systematic numerical evaluation of the $\curlyP$ factors \eqref{eq:curlyP} has proven to be a major challenge so far. 
%Indeed it is not clear yet whether the factorization scheme proposed in \cite{2017NJPh...19h3001B} is the best approach to evaluating the free generating functional. 
%We will see how to deal with this problem once we evaluate the analytic expressions.

%%%%%%%%%%%%%%%%%%%%%%%%%%%%
\begin{figure}[ht]
    \centering
    \includegraphics[scale = 0.8, clip = true, trim = 1cm 10cm 3cm 0cm]{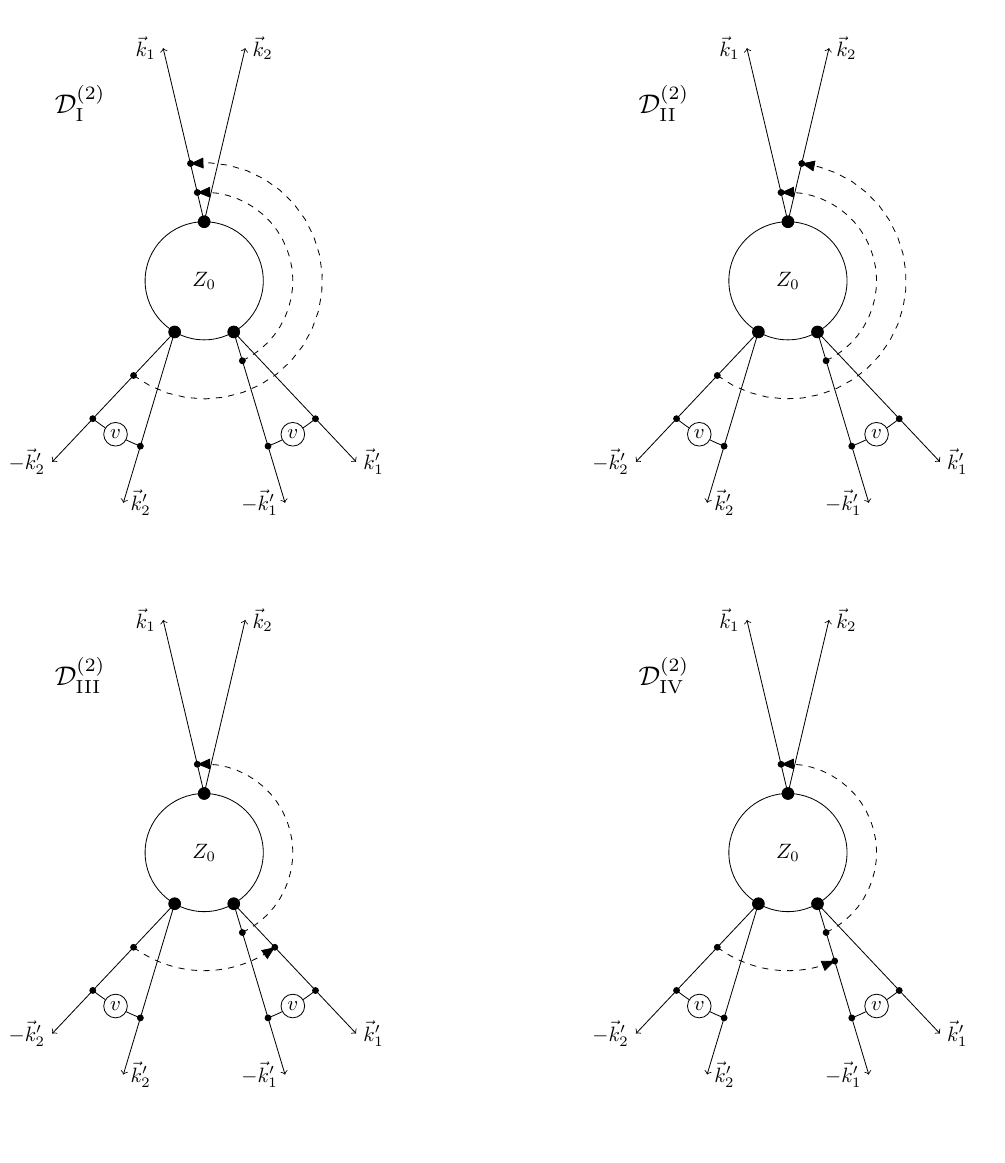}
    \caption{Diagrammatic representation of \eqref{eq:secondOrderContractions} according to the prescription in \cite{2017NJPh...19h3001B}. The corresponding multiplicity can be read off each diagram by permuting the external and internal wave-vectors and determining how many permutations result in a distinct configuration. For example, for diagram $\mathcal{D}^{(2)}_\mathrm{I}$ both internal wave-vectors connect to the same external wave-vector. Permuting the indices of the external wave-vectors leads to two distinct configurations, whereas a permutation of the internal wave-vectors has no such effect. The resulting multiplicity is therefore 2.}
    \label{fig:diagramsSecondOrder}
\end{figure}
%%%%%%%%%%%%%%%%%%%%%%%%%%%%
%%%%%%%%%%%%%%%%%%%%%%%%%%%%%%%%%%%%%%%%%%%%%%
\subsection{Second-order corrections}
Including the interaction operator at second order---term 3 in \eqref{eq:two-point function second order}---gives rise to a total of four distinct diagrams, which are shown in \autoref{fig:diagramsSecondOrder}. 
The four diagrams correspond to the distinct contractions of the response field operators
\begin{align}\label{eq:secondOrderContractions}
    &\tfrac{1}{2}(\ii\hat{S})^2_\mathcal{I} \ \hat{\rho}(1) \hat{\rho}(2) Z_0[\tensorJ, \tensorK] \Big|_{\tensorJ = \tensorK = 0} \nonumber
    \\[3mm]
    &=
    \frac{\ii^2}{2} \int \dd 2' v(2') \hat{B}(-2') \rho(2')
    \Big[2 \int \dd 1' \ v(1') \wick[offset = 1.2em]{\c1{\hat{B}}(-1') \hat{\rho}(1') \c1{\hat{\rho}}(1) \hat{\rho}(2)} \Big] Z_0[\tensorJ, \tensorK] \Big|_{0} \nonumber
    \\[3mm]
    &= \ii^2\int \dd 2' \int \dd 1' \ v(2') v(1') 
    \Big\{\wick{\c2{\hat{B}}(-2')\c1{\hat{B}}(-1') \hat{\rho}(2') \hat{\rho}(1') \c1{\hat{\rho}}\c2{\vphantom{\hat{\rho}}}(1) \hat{\rho}(2)} + 
    \wick{\c2{\hat{B}}(-2')\c1{\hat{B}}(-1')  \hat{\rho}(2') \hat{\rho}(1') \c1{\hat{\rho}}(1) \c2{\hat{\rho}} (2)} \nonumber
    \\[2mm]
    &\qquad \quad + 2
    \wick{\c2{\hat{B}}(-2')\c1{\hat{B}}(-1')  \hat{\rho}(2') \c2{\hat{\rho}}(1') \c1{\hat{\rho}}(1) \hat{\rho}(2)}
    + 2
    \wick{\c2{\hat{B}}(-2')\c2{\hat{B}}\c1{\vphantom{\hat{B}}}(-1')  \hat{\rho}(2') \hat{\rho}(1') \c1{\hat{\rho}}(1) \hat{\rho}(2)} \Big\} Z_0[\tensorJ, \tensorK] \Big|_{0} \nonumber
    \\[2mm]
    &\eqdef \mathcal{D}^{(2)}_\mathrm{I} (1, 2) + \mathcal{D}^{(2)}_\mathrm{II}(1, 2) + \mathcal{D}^{(2)}_\mathrm{III} (1, 2)+ \mathcal{D}^{(2)}_\mathrm{IV}(1, 2).
\end{align}
In the second line above we explicitly applied another interaction operator to the first-order result. 
The response field operator $\hat{B}(-2')$ has four non-vanishing possibilities to connect to other fields, resulting in the expression spreading over the third and fourth lines. 
The first and second terms, corresponding to the diagrams $ \mathcal{D}^{(2)}_\mathrm{I}$ and $\mathcal{D}^{(2)}_\mathrm{II}$, respectively, both acquire a multiplicity factor of 2, whereas the third and fourth terms $ \mathcal{D}^{(2)}_\mathrm{III}$ and $\mathcal{D}^{(2)}_\mathrm{IV}$ are multiplied by 4.
At this point writing out the resulting expressions becomes rather cumbersome, such that we will merely present the calculations for the diagram $\mathcal{D}^{(2)}_\mathrm{III}$, given by
\begin{align}\label{eq:diagram 23}
    \mathcal{D}^{(2)}_\mathrm{III}(1,2) = \ii^2
    \int \dd 2' \int \dd 1' \ v(2') v(1') b(1,-1') b(1',-2') G^{(0)}_{\rho\rho\rho\rho}(1-1', 2, 1'-2', 2').
\end{align}
At second order in the interaction operator we generate four shift vectors in the generating functional, the components of which are determined by the contractions of the response field operators, i.e.\ by the specific diagram we consider. They can be directly read off the arguments of the free four-point function in \eqref{eq:diagram 23}:
\begin{align*}
    \vecL_{q_1} &= - \veck_1 + \veck'_1, \qquad \qquad \vecL_{p_1} = - g_{qp}^{(\mathcal{Z})}(\eta_1,0) \veck_1 + g_{qp}^{(\mathcal{Z})}(\eta'_1,0) \veck'_1,
    \\
    \vecL_{q_2} &= - \veck_2, \qquad \qquad \qquad \, \vecL_{p_2} = - g_{qp}^{(\mathcal{Z})}(\eta_1,0) \veck_2,
    \\
    \vecL_{q_3} &= - \veck'_1 + \veck'_2, \qquad \qquad   \vecL_{p_3} = - g_{qp}^{(\mathcal{Z})}(\eta'_1,0) \veck'_1 + g_{qp}^{(\mathcal{Z})}(\eta'_2,0) \veck'_2,
    \\
    \vecL_{q_4} &= - \veck'_2, \qquad \qquad \qquad \,  \vecL_{p_4} = - g_{qp}^{(\mathcal{Z})}(\eta'_2,0) \veck'_2.
\end{align*}
Factorizing the resulting free generating functional and carrying out the trivial integrals over auxiliary wave vectors yields
\begin{multline}\label{eq:generating functional second order}
    G^{(0)}_{\rho\rho\rho\rho}(1-1', 2, 1'-2', 2')=
    (2\pi)^3 \Bar{\rho}^{\,4} 
    \delta_D\big(\veck_1+ \veck_2 \big)
    \ee^{-{Q}_D(\tensorL_p)} 
    \Big[
    (2\pi)^3\delta_D(\vec{k}_1 - \vec{k}'_1) (2\pi)^3\delta_D(\vec{k}_1 - \vec{k}'_2) \mathcal{P}_{42}(\vec{k}_1)\\[2mm]
    +(2\pi)^3\delta_D(\vec{k}'_1 - \vec{k}'_2) \mathcal{P}_{21}(\vec{k}_1)\mathcal{P}_{41}(\vec{k}'_1)
    +(2\pi)^3\delta_D(\vec{k}'_1 - \vec{k}'_2) \mathcal{P}_{21}(\vec{k}_1 - \vec{k}'_1)\mathcal{P}_{42}(\vec{k}'_1)\\[2mm]
    +(2\pi)^3\delta_D(\vec{k}_1 - \vec{k}'_2) \mathcal{P}_{31}(\vec{k}_1 - \vec{k}'_1)\mathcal{P}_{42}(\vec{k}_1)
    +(2\pi)^3\delta_D(\vec{k}_1 - \vec{k}'_1 + \vec{k}'_2) \mathcal{P}_{41}(\vec{k}_1 - \vec{k}'_1)\mathcal{P}_{32}(\vec{k}_1)\\[2mm]
    +(2\pi)^3\delta_D(\vec{k}'_1 - \vec{k}'_2) \mathcal{P}_{41}(\vec{k}_1 - \vec{k}'_1)\mathcal{P}_{42}(\vec{k}_1)
    +(2\pi)^3\delta_D(\vec{k}_1 - \vec{k}'_1) \mathcal{P}_{32}(\vec{k}_1 - \vec{k}'_2)\mathcal{P}_{42}(\vec{k}'_2)\\[2mm]
    +(2\pi)^3\delta_D(\vec{k}_1 - \vec{k}'_1) \mathcal{P}_{32}(\vec{k}_1)\mathcal{P}_{43}(\vec{k}'_2)
    +(2\pi)^3\delta_D(\vec{k}_1 - \vec{k}'_1) \mathcal{P}_{42}(\vec{k}_1)\mathcal{P}_{43}(\vec{k}_1 - \vec{k}'_2) +\mathcal{O}(\curlyP^3)\Big].
\end{multline}
We choose to show the expression only up to second order in $\curlyP$, but in fact the generating functional at second order perturbation theory contains terms of up to sixth order in the $\curlyP$ factors. 
{
In general, the factorized expression for a free $n$-point correlation function has a total of $2^{N_\mathrm{aux}}$ terms, where $N_\mathrm{aux} = \frac{n(n-1)}{2}$ is the number of auxiliary wave-vectors.
The maximum number of $\curlyP$ factors is given by $N_\mathrm{aux}$.
The expressions of the generating functional contributions for the other three diagrams at second order in $\curlyP$ can be found in \autoref{apx:diagrams}.
We will see in the following that, at our current level of the analysis, there is no need for us to include higher orders in $\curlyP$.}

\section{Numerical evaluation}
\label{sec:numerical results}
% motivate expansion of curly P
As stated previously, the expressions \eqref{eq:first order contraction} and \eqref{eq:secondOrderContractions} together with the corresponding expressions for the generating functional contain the full information on initial phase-space correlations of the classical particles.
These are contained in the $\curlyP$ factors which, in their most general form, are given by \eqref{eq:curlyP}
and depend on the initial momentum covariance matrix $C_{pp}(q)$. 
In the special case where both momentum shift vectors are anti-parallel and align with $\veck$, i.e.\ $\vecL_{p_i} \parallel -\vecL_{p_j}$ and $\vecL_{p_i} \parallel \veck$ we recover the Zel'dovich power spectrum in our analysis.
General expressions in KPT do not fulfill these conditions; in fact there is no restriction on the length and orientation of the shift vectors w.r.t.\ each other. 
In such a general case the integral in \eqref{eq:curlyP} is very challenging to solve numerically and is yet to be evaluated in a reliable way.
As a step towards that goal recent progress has been made on the asymptotic behaviour of \eqref{eq:curlyP} \cite{2021-2110.07427_Konrad, 2022-2202.08059_Konrad}, which can be used to cover especially challenging parts of the parameter space where rapid oscillation of the integration kernel make a numerical evaluation impossible. 
However, more work is needed to achieve a stable integration of such expressions in general, such that we have to resort to an expansion of \eqref{eq:curlyP} in terms of the initial momentum covariance matrix.

By the definition \eqref{eq:Cpp in terms of power spectrum} of $C_{pp}(q)$ in terms of the initial density fluctuation power spectrum, expanding the exponent in \eqref{eq:curlyP} results in 
\begin{align}
     \curlyP_{ij}(\veck) &= \curlyP_{ij}^{(1)}(\veck) + \curlyP_{ij}^{(2)}(\veck) + \mathcal{O}\Big(\big(P^{(\ii)}_\delta\big)^3\Big)
     \nonumber
     \\[2mm]
     \mathrm{with} \quad
     \curlyP_{ij}^{(1)}(\veck)  &= -\frac{(\vecL_{p_i} \cdot \veck) (\vecL_{p_j} \cdot \veck)}{\veck^{4}} P^{(\ii)}_\delta (k) \label{eq:curlyP-1}
     \\
     \curlyP_{ij}^{(2)}(\veck)  &= \frac{1}{2} \int_{k'} \frac{(\vecL_{p_i} \cdot \veck') (\vecL_{p_j} \cdot \veck')}{\veck^{\prime 4}}
     \frac{\big[\vecL_{p_i} \cdot (\veck'-\veck)\big] \big[\vecL_{p_j} \cdot (\veck'-\veck)\big]}{(\veck'-\veck)^{4}}
     P^{(\ii)}_\delta (k') P^{(\ii)}_\delta (|\veck' - \veck|) \label{eq:curlyP-2}. 
\end{align}
% motivate why we stop at 1-loop
In our analysis we choose to truncate the expansion at second order in the initial power spectrum, i.e.\ at one-loop order. 
Going beyond one-loop order is not a substantial conceptual step, but entails a considerable amount of additional calculations that we want to postpone here.
The reason for this is two-fold: 
(i) The main result of this work is to illustrate and compare the effects of first and second order KPT and not of higher-order loop expansions.
(ii) Since a loop expansion is not intrinsically necessary for KFT we regard this result as a stepping stone to a consistent evaluation that eventually relies on a numerical evaluation of the full $\curlyP$ factors.
Nevertheless, including higher loop orders should be part of future investigations.

Considering eq. \eqref{eq:generating functional second order} for the generating functional at second-order KPT, it becomes clear that going beyond second order in $\curlyP$ is not necessary in our analysis, because such terms will contribute only beyond one-loop order.
% motivate why we leave aside tree level terms in the interaction
As we have seen in \autoref{sec:particle trajectories}, choosing Zel'dovich trajectories as inertial motion changes the two-particle interaction potential to a form that can be modelled by a Yukawa potential. 
From the exact expression of the collective interaction potential \eqref{eq:collectivePotential} we know that only strictly non-linear density contrasts can be the source of interactions.
However, due to the imperfect approximation of the two-particle interaction potential by a Yukawa potential, contributions from the linear power spectrum in the interaction term are still present, albeit suppressed. 
Since we know that such contributions should not appear, we neglect all tree-level contributions in the interaction terms. 
In terms of the initial power spectrum the lowest-order contribution to the freely evolving power spectrum is thus the linear power spectrum, whereas corrections due to particle interactions should only contribute at the one-loop level.

%From the fact that the collective potential is sourced exclusively by density contrasts that go beyond the linear density contrast we draw the following consequences:
%(i) We may approximate the result of free KFT by a linear power spectrum, since interactions modify the density contrast such that it deviates from this result.
%(ii) We may neglect terms in the interaction that are linear in the initial power spectrum and instead concentrate on the one-loop expression of the KFT perturbation series.

% try to motivate why we approximate the damping by a fraction
The Gaussian damping factor needs a separate treatment. 
In common loop expansions in SPT and LPT the fundamental expansion parameter is the initial power spectrum, and consequently also the momentum dispersion in \eqref{eq:damping}.
In such a framework it is therefore important to compare terms of the same combined order in $\sigma_p^2$ and $P_\delta^{(\ii)}$.
In KFT the actual perturbative expansion takes place in the interaction operator, i.e.\ in deviations of particle trajectories from inertial motion.
A loop expansion is not intrinsically part of the theory and merely provides a temporary tool to allow for a numerical evaluation of \eqref{eq:generating functional first order} and \eqref{eq:generating functional second order}
It is thus bound to fail if the damping factor is included in the usual way, as suggested by \cite{2015_PhysRevD.91.103507}.
In order to approximate the effect of the exponential damping for the perturbative correction terms at one-loop order we replace it by $\frac{1}{1+Q_D}$, following \cite{2016NJPh...18d3020B}. 
%As was shown in \cite{2021-2110.07427_Konrad} the Gaussian damping factor is cancelled exactly by the first order contribution of $\curlyP$ in the asymptotic regime for a free power spectrum.
%Based on this premise it is therefore natural to assume that the same is true for the 
This approximation ensures that we retain the essential properties of the exponential damping: (i) It stays positive for arbitrary values of $k$, but (ii) introduces a damping effect on small scales that (iii) is weakened to account for the smaller contribution of correlations due to the one-loop expansion of the $\curlyP$ factors.
Using this approximation, large negative contributions do not occur in our correction terms.
Note that this treatment of the exponential damping is only applied to the one-loop correction terms and not to the free result which we approximate by the linear power spectrum, i.e.\ we evaluate at tree-level.
Using \eqref{eq:curlyP-1} and \eqref{eq:curlyP-2}, we thus approximate the three- and four-point functions \eqref{eq:generating functional first order} and \eqref{eq:generating functional second order} corresponding to the correction terms due to particle interactions at first and second order, respectively, by
\begin{multline}\label{eq:first order pt one-loop}
   G^{(0)}_{\rho\rho\rho}(1-1', 2, 1') \approx
    (2\pi)^3 \Bar{\rho}^{3} 
    \delta_D\left(\veck_1+ \veck_2 \right)
    \frac{1}{1+{Q}_D(\tensorL_p)} 
    \Big[
   (2\pi)^3 \delta_D(\veck_1'-\veck_1)\, \curlyP^{(2)}_{32}(\veck_1)
   \\[2mm]
   +
   \curlyP^{(1)}_{21}(\veck_1)\,\curlyP^{(1)}_{31}(\veck_1')
   +
   \curlyP^{(1)}_{21}(\veck_1'-\veck_1)\,\curlyP^{(1)}_{32}(\veck'_1)
   +
   \curlyP^{(1)}_{31}(\veck_1'-\veck_1)\,\curlyP^{(1)}_{32}(\veck_1)
    \Big].
\end{multline}
\begin{multline}\label{eq:second order pt one-loop}
    G^{(0)}_{\rho\rho\rho\rho}(1-1', 2, 1'-2', 2') \approx
    (2\pi)^3 \Bar{\rho}^{4} 
    \delta_D\big(\veck_1+ \veck_2 \big)
    \frac{1}{1+{Q}_D(\tensorL_p)}\\[2mm]
    \Big[
    (2\pi)^3\delta_D(\vec{k}_1 - \vec{k}'_1) (2\pi)^3\delta_D(\vec{k}_1 - \vec{k}'_2) \mathcal{P}^{(2)}_{42}(\vec{k}_1)
    +(2\pi)^3\delta_D(\vec{k}'_1 - \vec{k}'_2) \mathcal{P}^{(1)}_{21}(\vec{k}_1)\mathcal{P}^{(1)}_{41}(\vec{k}'_1)\\[2mm]
    +(2\pi)^3\delta_D(\vec{k}'_1 - \vec{k}'_2) \mathcal{P}^{(1)}_{21}(\vec{k}_1 - \vec{k}'_1)\mathcal{P}^{(1)}_{42}(\vec{k}'_1)
    +(2\pi)^3\delta_D(\vec{k}_1 - \vec{k}'_2) \mathcal{P}^{(1)}_{31}(\vec{k}_1 - \vec{k}'_1)\mathcal{P}^{(1)}_{42}(\vec{k}_1)\\[2mm]
    +(2\pi)^3\delta_D(\vec{k}_1 - \vec{k}'_1 + \vec{k}'_2) \mathcal{P}^{(1)}_{41}(\vec{k}_1 - \vec{k}'_1)\mathcal{P}^{(1)}_{32}(\vec{k}_1)
    +(2\pi)^3\delta_D(\vec{k}'_1 - \vec{k}'_2) \mathcal{P}^{(1)}_{41}(\vec{k}_1 - \vec{k}'_1)\mathcal{P}^{(1)}_{42}(\vec{k}_1)\\[2mm]
    +(2\pi)^3\delta_D(\vec{k}_1 - \vec{k}'_1) \mathcal{P}^{(1)}_{32}(\vec{k}_1 - \vec{k}'_2)\mathcal{P}^{(1)}_{42}(\vec{k}'_2)
    +(2\pi)^3\delta_D(\vec{k}_1 - \vec{k}'_1) \mathcal{P}^{(1)}_{32}(\vec{k}_1)\mathcal{P}^{(1)}_{43}(\vec{k}'_2)\\[2mm]
    +(2\pi)^3\delta_D(\vec{k}_1 - \vec{k}'_1) \mathcal{P}^{(1)}_{42}(\vec{k}_1)\mathcal{P}^{(1)}_{43}(\vec{k}_1 - \vec{k}'_2) \Big].
\end{multline}
The three remaining diagrams of the second order expression \eqref{eq:secondOrderContractions} are evaluated in an equivalent fashion.
To arrive at the numerical results presented in the following, we have to integrate expressions like \eqref{eq:first order pt one-loop} and \eqref{eq:second order pt one-loop}.
Since these involve multidimensional integrals that have to be solved numerically, we are using Monte Carlo integration methods from the Gnu Scientific Library to evaluate the expressions.

%%%%%%%%%%%%%%%%%%%%%%%%%%%%%%%%%%%%%%%%%%%%%%%%%%%%%%%%%%
\subsection{Results}
% give details on the numerical evaluation
For the numerical evaluation we choose a universe with dark matter density $\Omega_c h^2=0.122$, baryon density $\Omega_b h^2=0.022$, $\sigma_8=0.8$, Hubble parameter $h=0.675$ and spectral index $n_s =0.965$.
%For all calculations the initial and final redshifts are respectively $1100$ and $0$ unless indicated otherwise.
The initial power spectrum is generated using the CAMB library \cite{2011ascl.soft02026L_CAMB}. 
The non-linear power spectrum from KFT is compared to the emulated simulation results of the Cosmic Emu \cite{2016ApJ...820..108H_emu}.
%%%%%%%%%%%%%%%%%%%%%%%%%%%%%%%%%%%%%%%%%%%%%%
\begin{figure}[ht]
    \centering
    \includegraphics[scale = 0.47]{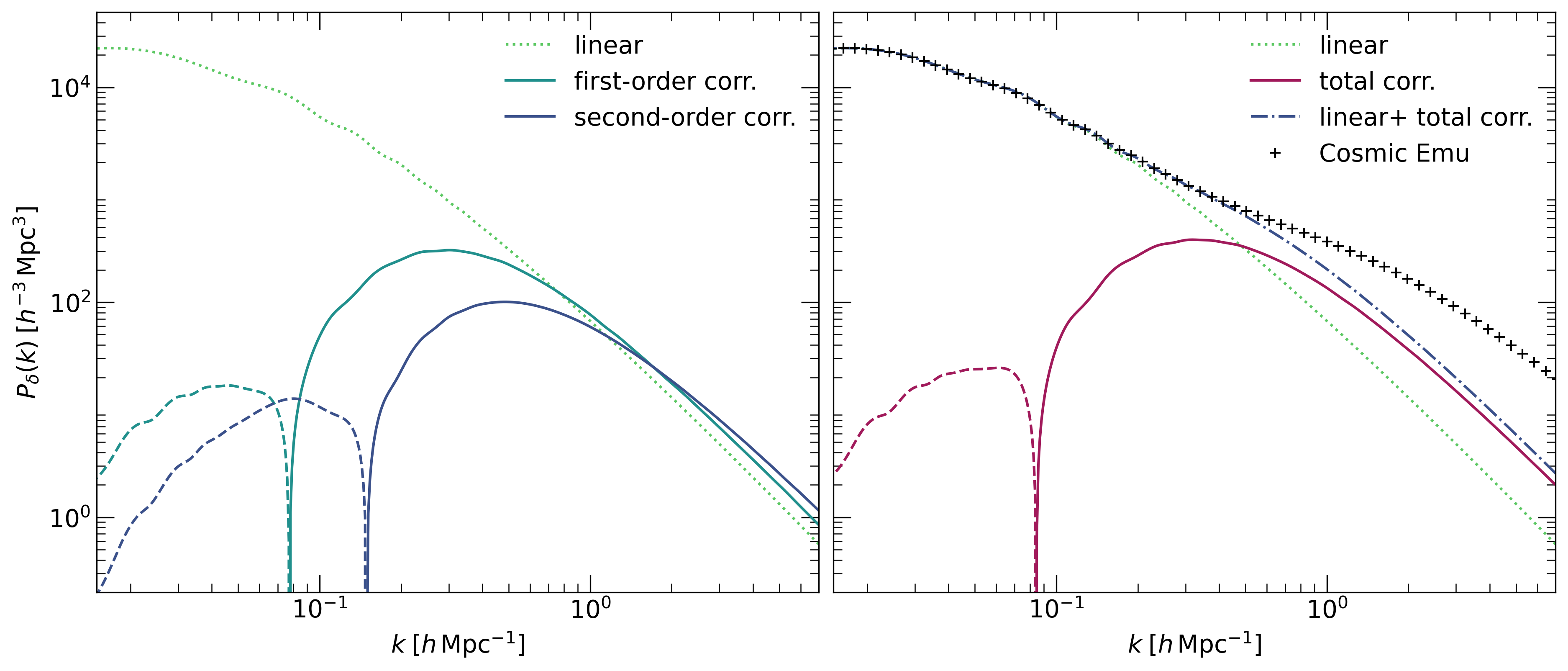}
    \caption{\textbf{Left panel:} Corrections to the linear CAMB power spectrum (light green, dotted line) from first-oder (green) and second-order (blue) KPT. The two contributions from perturbation theory show a similar behaviour, with the second-order result being shifted towards larger $k$ and and smaller amplitude. Preliminary results on higher orders show that this is a systematic behaviour which continues as we go towards higher orders.
    \textbf{Right panel:} The full contribution of first- and second-order corrections (crimson) and the resulting non-linear power spectrum (blue dash-dotted) are shown. Additionally, the emulated simulation result from the Cosmic Emu (plus symbols) \cite{2016ApJ...820..108H_emu} and the linear CAMB power spectrum (light green, dotted) are plotted for reference. 
    %The free parameter $\kappa$ of the Yukawa cutoff \eqref{eq:Yukawa cutoff} is set to $0.85$.
    }
    \label{fig:fig1 first and second order corrections}
\end{figure}
%%%%%%%%%%%%%%%%%%%%%%%%%%%%%%%%%%%%%%%%%%%%%%
In the left panel of \autoref{fig:fig1 first and second order corrections} we show the result for the correction terms to the linear power spectrum due to first- and second-order interactions.

There are a couple of points to make about the first-order result, which has been derived earlier in \cite{2016NJPh...18d3020B}.
Firstly, in contrast to previous calculations, which used a modified version of Zel'dovich trajectories, we work with standard Zel'dovich trajectories to describe the inertial motion of dark matter particles. 
Secondly, we have shown in \autoref{apx:trajectories} that the collective interaction potential for particles on inertial Zel'dovich trajectories fulfills a modified Poisson equation and that the resulting two-particle interactions can be modelled by a Yukawa potential. 
Both modifications lead to an improvement in the behaviour of the first-order result compared to \cite{2016NJPh...18d3020B}, such that the transition from the linear to the non-linear power spectrum matches results from simulations more accurately.

The correction due to second-order perturbations shows a similar behaviour to the first-order result: On large scales we recover a negative contribution, leading to a stronger (albeit still weak) suppression of the power spectrum, whereas structures on intermediate to small scales are further enhanced.
The shift of the maximum of the correction term towards smaller scales (compared to the first-order term) can be qualitatively understood by the increasingly important role interactions between multiple particles play as we move towards smaller scales.
Preliminary results on perturbations beyond second order show that this behaviour is continued for higher orders and will be the topic of future work.

The total correction up to second-order in KPT, as well as the resulting non-linear power spectrum are shown in the right panel of \autoref{fig:fig1 first and second order corrections}. 
As mentioned above it shows that the transition from the linear to the non-linear bump is captured more accurately compared to earlier works on KFT \cite{2016NJPh...18d3020B}, which can be attributed to the modified two-particle interaction potential, i.e. the time-dependent Yukawa cutoff reducing the contribution of interactions on intermediate to large scales.
On small scales the solution falls below the result from the cosmic emulator, which is expected for two reasons:
(i) On small scales particle interactions become increasingly important, such that we need to go to higher orders, or beyond perturbation theory altogether in order to capture the full non-linear growth of structures. One promising alternative to perturbation theory is to employ a mean-field approach \cite{2020arXiv_MeanField} to approximate the interaction operator.
(ii) On small scales density contrasts become large, such that the expansion of the Gaussian damping and of the $\curlyP$ factors in orders of the initial power spectrum breaks down. This will need to be addressed by further work on the numerical evaluation of the $\curlyP$ factors.
%%%%%%%%%%%%%%%%%%%%%%%%%%%%%%%%%%%%
\begin{figure}[ht]
    \centering
    \includegraphics[scale = 0.47]{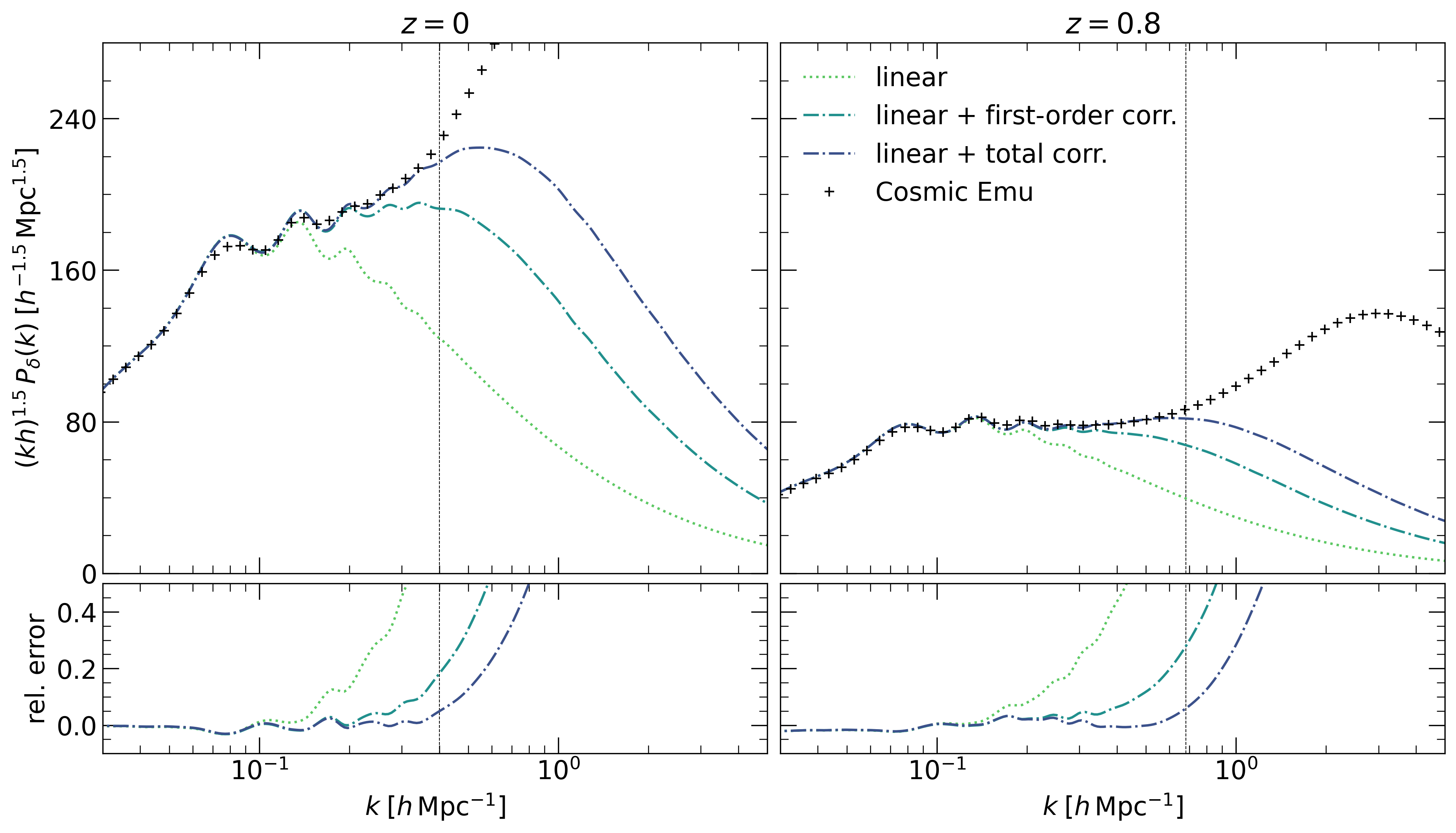}
    \caption{Comparison of the non-linear power spectra from up to first-order (green dash-dotted) and up to second-order (blue dash-dotted) KPT for $z=0$ (left panel) and $z=0.8$ (right panel). The linear CAMB (light green dotted) and the Cosmic Emu (plus symbols) power spectra are shown for comparison. 
    The lower panels show the relative deviation of the CAMB and KFT results w.r.t.\ the non-linear result from the Cosmic Emu.
    One can observe the systematic improvement towards smaller scales when going from first- to second-order perturbation theory.
    The vertical line marks the wavenumber $k$ below which the mismatch between emulated simulation and second-order KFT is smaller than $5\%$. 
    %We fixed $\kappa$ at $0.85$ and $0.7$ for $z=0$ and $z=0.8$, respectively.
    }
    \label{fig:fig3 comparison at different z}
\end{figure}
%%%%%%%%%%%%%%%%%%%%%%%%%%%%%%%%%%%%

\autoref{fig:fig3 comparison at different z} shows the results for the non-linear power spectrum for the two redshifts $z=0$ and $z=0.8$, as well as the relative difference to the simulation results.
To reduce the vertical range we plotted the product $(kh)^{1.5} P_\delta(k)$.
The plot nicely illustrates the improvement that second order corrections add to the non-linear KFT power spectrum.
Going to first order in KPT leads to a good agreement with simulation results until $k\approx 0.3\ h \, \mathrm{Mpc}^{-1}$ at redshift 0. On smaller scales, the agreement quickly deteriorates.
Including corrections up to second-order has little to no effect on scales $k<0.3 \ h \,\mathrm{Mpc}^{-1}$, but takes over on smaller scales where first-order effects are not sufficient anymore. 
Thus, the region where the deviation from the simulated power spectrum is below $5\%$ is extended to $k \approx 0.4\ h\, \mathrm{Mpc}^{-1}$ at redshift 0. The boundary of this region is marked by the dotted vertical line.
As mentioned above, preliminary results of higher-order corrections show that this trend continues, such that perturbations of higher orders progressively add more power on smaller scales, leaving intermediate scales untouched.
%We are aware that the amplitude of the BAO peaks is too large in the non-linear KFT power spectrum, which we trace back to the approximation of the damping factor. 
%Since a precise analysis of BAO's in the non-linear power spectrum is not the goal of this paper we did not split the power spectrum into a wiggly and a non-wiggly part as suggested by 
Reassuringly, the behaviour that we observe for results at redshift 0 is recovered for a large range of redshifts, as illustrated by the right panel. 
At redshift $0.8$, non-linear growth is limited to smaller scales than for $z=0$, such that the region of sub-$5\%$ deviations extends up to $k \approx 0.7\ h \,\mathrm{Mpc}^{-1}$ for the second-order result.

\section{Conclusion}
Building on previous work on KFT for cosmic structure formation \cite{2016NJPh...18d3020B, 2017NJPh...19h3001B}, we derived results for second-order perturbation theory in KFT. 
We started with a general introduction into the basic formalism of the theory and showed different means to calculate the effect of the interaction operator.
By linking contractions of response fields to density fields introduced in \cite{2015PhRvE..91f2120V} to the diagrammatic approach from \cite{2017NJPh...19h3001B}, we showed how the action of the interaction operator on a free correlation function leads to expressions in terms of higher-order correlations.
To recover linear growth on large scales in our free solution we introduced Zel'dovich trajectories and showed that this naturally leads to a modification of the collective potential, with the resulting two-particle interaction potential being of Yukawa-like shape.

We then derived analytical expressions for first- and second-order corrections in KFT in terms of the factorized generating functional.
The resulting expressions \eqref{eq:generating functional first order} and \eqref{eq:generating functional second order} contain momentum correlations up to infinite order through the factors $\curlyP$ of the generating functional.
They are natural corrections to the Zel'dovich power spectrum, which is the non-interacting two-point function in KFT. 
%Until this point no loop expansion has been made and perturbations purely take the form of deviations of particle motion from Zel'dovich trajectories.
%The resulting corrections to the linear power spectrum due to an expansion of the interaction operator have then been derived and evaluated using Monte Carlo integration methods.
Since a numerical evaluation of the full factors $\curlyP$ of the generating functional is so far out of reach we employed an expansion of these factors in terms of the initial momentum covariance matrix.
The main results of the numerical evaluation can be seen in \autoref{fig:fig1 first and second order corrections} and \autoref{fig:fig3 comparison at different z}, which show the result of including particle interactions up to second order in KFT.
The similarity of the contributions of first- and second-order is quite remarkable and does not trivially follow from the analytical expressions of the respective orders.
This similarity certainly calls for further investigation. 
The two main aspects of the curves that we see in \autoref{fig:fig1 first and second order corrections} are the fact that the second-order contribution (i) is shifted towards smaller scales and (ii) has a smaller amplitude.
Both properties are consistent with the intuition that particle interactions play a more pronounce role as we consider structures on smaller, more non-linear scales.
Compared to the first-order results, the second-order extends the agreement of the non-linear power spectrum with simulation results towards larger $k$. Deviations remain below $5\%$ up to $k \approx 0.4\ h \,\mathrm{Mpc}^{-1}$ for $z=0$ and $k\approx 0.7\ h \,\mathrm{Mpc}^{-1}$ for $z=0.8$.

{While treating the evolution cosmic structures with a particle-based approach has formal advantages, some tasks can be complicated by this framework.
Since we are not directly working with the dynamics of macroscopic fields, the treatment of biased tracers, such as dark matter halos, could be rather tricky in KFT.
One approach could consist in looking for maxima in the density field and again predicting their dynamics with KFT.}

Two main issues remain to be solved in the future, the first and more pressing challenge being the numerical evaluation of the factors $\curlyP$ of the generating functional.
The main strength of KFT lies in the fact that, in principle, it does not rely on a loop expansion, which necessarily breaks down on small scales.
Therefore, evaluating late-time correlation functions in terms of the full hierarchy of initial momentum correlations will be crucial.
Secondly, the relation and behaviour of higher-order perturbations needs to be studied in more detail, which will be the topic of future work.

\acknowledgments
We want to thank Robert Lilow, Elena Kozlikin, Tristan Daus and Iris Feldt for many helpful comments and interesting discussions.
This work was supported by the Luxembourg National Research Fund (13511583) and by the Deutsche Forschungsgemeinschaft (DFG, German Research Foundation) under Germany's Excellence Strategy EXC 2181/1 - 390900948 (the Heidelberg STRUCTURES Excellence Cluster).

\appendix
\section{Derivation of particle trajectories}\label{apx:trajectories}

We start with the Lagrangian for classical point-like particles in an expanding universe with the time coordinate introduced in \eqref{eq:eta definition},

\begin{align}\label{eq:LagrangianEta}
    L\bigg(\vec{q},\frac{\dd \vec{q}}{\dd \eta},\eta\bigg) = \frac{1}{2}ma^2\bigg(\frac{\dd \vec{q}}{\dd \eta}\bigg)^2Hf-\frac{m\phi(\vec{q})}{Hf}, 
\end{align}
where $H$ is the Hubble function and $f:=\frac{\dd \ln{D_+}}{\dd \ln{a}}$. The potential satisfies the Poisson equation
\begin{align}
    \vec{\nabla}^2 \phi = 4\pi G a^2(\rho_m(q)-\Bar{\rho}_m),
\end{align}
with the mass density function $\rho_m(q)$ and the average density $\Bar{\rho}_m$. The next step is to construct the Hamiltonian of the system
\begin{align}
   \mathcal{H}(\vec{p}_\mathrm{c},\vec{q},\eta)=\frac{1}{2} \frac{{\vec{p}_{\mathrm{c}}}^{\;2}}{a^2Hf} +\frac{\phi}{Hf},
\end{align}
in order to obtain the equations of motion for the point-like particles in an expanding universe. The subscript ``c" stands for canonical. With the help of Hamilton's equations we arrive at two first-order ordinary differential equations which describe the phase-space dynamics
\begin{align}\label{eq:Hamilton equations}
\frac{\dd \vec{q}}{\dd \eta} &= \vec{p} \nonumber\\
\frac{\dd \vec{p}}{\dd \eta} &\approx -\frac{1}{2}\vec{p} -\vec{\nabla}\tilde{\phi}.
\end{align}
Here we have introduced new definitions for particle momentum and interaction potential in \eqref{eq:Hamilton equations} such that $\vec{p}=\frac{\vec{p}_\mathrm{c}}{a^2mHf}$ and $\tilde{\phi}=\frac{\phi}{a^2H^2f^2}$. 
We also used $\frac{\Omega_\mathrm{m}}{f^2} \approx 1$, which is a good approximation in the matter- and early $\Lambda$-dominated epochs of a $\Lambda$CDM cosmology \cite{BERNARDEAU20021}. The new potential satisfies the Poisson equation
\begin{align}\label{eq:PoissonTilde}
    \nabla^2\tilde{\phi} &= \frac{4\pi G}{H^2f^2} \Omega_\mathrm{m}\rho_\mathrm{crit}\delta(\vec{q},\eta) \nonumber\\
    &=\frac{4\pi G}{H^2f^2} \frac{3H^2\Omega_\mathrm{m}}{8\pi G}\delta(\vec{q},\eta) \nonumber\\
    &\approx\frac{3}{2}\delta(q,\eta).
\end{align}
We can write \eqref{eq:Hamilton equations} as a matrix equation by collecting the phase-space coordinates in $\vec{x}=(\vec{q},\vec{p})^{\mathrm{T}}$,
\begin{align}\label{eq:Hamilton eq. combined}
    \frac{\dd}{\dd\eta}\vec{x}+\mathcal{A}_6\vec{x}=\vec{\mathcal{F}}(\bar{\eta}) \quad \text{where} \quad \mathcal{A}_6 = \begin{pmatrix} 0_3 & -\mathcal{I}_3 \\ 0_3 & \frac{1}{2}\mathcal{I}_3\end{pmatrix}; \quad \vec{\mathcal{F}}(\bar{\eta}) = \begin{pmatrix} 0 \\ -\vec{\nabla}\tilde{\phi} \end{pmatrix}.
\end{align}
 The Green's function for this differential equation reads
\begin{align}
    \mathcal{G}(\eta,\eta')=& \exp{[-\mathcal{A}_6(\eta-\eta')]}\Theta(\eta-\eta') \nonumber\\
    =& \mathcal{I}_6 + \sum_{n=1}^\infty \frac{(-1)^n}{n!}[(\eta-\eta')\mathcal{A}_6]^n \Theta(\eta-\eta')\nonumber\\
    =& \mathcal{I}_6 + \begin{pmatrix} 0_3 & -\mathcal{I}_3\sum_{n=1}^\infty \frac{(-1)^n}{n!}(\eta-\eta')^n 2^{-n+1} \\ 0_3 & \mathcal{I}_3\sum_{n=1}^\infty \frac{(-1)^n}{n!}(\eta-\eta')^n 2^{-n}\end{pmatrix}\Theta(\eta-\eta').
\end{align}
We compute the propagator as follows:
\begin{align}
  g_{qp}(\eta,\eta')=  \sum_{n=1}^\infty \frac{(-1)^{n+1}}{2^{n-1}n!}(\eta-\eta')^n = -2\sum_{n=1}^\infty \frac{(-1)^{n}}{2^{n}n!}(\eta-\eta')^n = 2\left(1-e^{-\frac{1}{2}(\eta-\eta')}\right),
\end{align}
and similarly we arrive at $g_{pp}(\eta,\eta')=  e^{-\frac{1}{2}(\eta-\eta')}$. Finally we can write the full phase-space trajectories as in \eqref{eq:particle trajectories Physical}. The spatial trajectories are also represented as
\begin{align}
    \vec{q}(\eta) = \vec{q}^{\,(\mathrm{i})} + g_{qp}(\eta,0)\vec{p}^{\,(\mathrm{i})} - \int_0^\eta \dd\eta'g_{qp}(\eta,\eta')\vec{\nabla}\tilde{\phi}.
\end{align}
\subsection{Free motion with Zel'dovich trajectories}
Since we like to capture linear structure growth within the free theory, we impose Zel'dovich trajectories as the inertial trajectory. For this, the differential equation \eqref{eq:Hamilton eq. combined} needs some adjustment. We rewrite \eqref{eq:Hamilton eq. combined} for the free case as

\begin{align}\label{eq:Differential Zeldovich}
     \frac{\dd}{\dd\eta}\vec{x}+A_6\vec{x}=0 \quad \text{where} \quad A_6 = \begin{pmatrix} 0_3 & \mathcal{I}_3 \\ 0_3 & \mathcal{I}_3 \end{pmatrix}  .
\end{align}
The Green's function for \eqref{eq:Differential Zeldovich} can be written as in \eqref{eq:Greens function Zeldovich} according to the steps applied for Newtonian trajectories. Imposing Zel'dovich trajectories modifies the inertial motion of the classical particles. In order to compensate for this modification and ensure that the full trajectories are unchanged, the interaction potential needs to be modified.
To compute the required modification we equate \eqref{eq:particle trajectories Physical} and \eqref{eq:particleTrajectories Zeldovich} as follows:
\begin{align}\label{eq:Modified Newtonian}
     \vec{q}(\eta) &= \vec{q}^{\,(\mathrm{i})} + g_{qp}(\eta,0)\vec{p}^{\,(\mathrm{i})} - \int_0^\eta \dd\eta'g_{qp}(\eta,\eta')\vec{\nabla}\tilde{\phi}. \nonumber \\
   \vec{q}(\eta) &= \vec{q}^{\,(\mathrm{i})} +g_{qp}^{(\mathcal{Z})}\vec{p}^{\,(\mathrm{i})}-\left[\int_0^\eta \dd\eta'g_{qp}(\eta,\eta')\vec{\nabla}\tilde{\phi} +(g_{qp}^{(\mathcal{Z})}-g_{qp}(\eta,0)) \vec{p}^{\,(\mathrm{i})}     \right].
\end{align}
Particle trajectories with free Zel'dovich trajectories and a time-dependent modification to the potential
\begin{align}\label{eq:ZEld.traj}
  \vec{q}(\eta) &= \vec{q}^{\,(\mathrm{i})} +g_{qp}^{(\mathcal{Z})}\vec{p}^{\,(\mathrm{i})} - \int_0^\eta \dd\eta'g_{qp}(\eta,\eta')(\vec{\nabla}\tilde{\phi}+A_p(\eta')\vec{p}^{\,(\mathrm{i})})
\end{align}
must represent the same physical trajectories. The equivalence of \eqref{eq:Modified Newtonian} and \eqref{eq:ZEld.traj} indicates the relation
\begin{align*}
    \int_0^\eta \dd\eta'g_{qp}(\eta,\eta') A_p(\eta') &= g_{qp}^{(\mathcal{Z})}-g_{qp}(\eta,0)\\
   \xrightarrow{\frac{\dd}{\dd\eta}} \int_0^\eta \dd\eta' e^{-\frac{1}{2}\eta}e^{\frac{1}{2}\eta'}A_p(\eta') &= e^\eta -  e^{-\frac{1}{2}\eta}  \\
    \int_0^\eta \dd\eta'e^{\frac{1}{2}\eta'}A_p(\eta') &= e^{\frac{3}{2}\eta}-1 \\
    A_p(\eta) &= \frac{3}{2}e^\eta .
\end{align*}
With this modification and using the linearised continuity equation to write $\vec{\nabla} \cdot \vec{p}^{\,(\mathrm{i})} \approx -\delta^ {(\mathrm{i})}$, one can show that the modified potential is satisfying the Poisson equation given in \eqref{eq:mod Poisson_eq}.

%%%%%%%%%%%%%%%%%%%%%%%%%%%%%%%%%%%%%%%%%%%%%%%%%%%%%%%
%%%%%%%%%%%%%%%%%%%%%%%%%%%%%%%%%%%%%%%%%%%%%%%%%%%%%%%
\section{Derivation of the two-particle interaction potential} \label{apx:Yukawa cutoff}
\begin{figure}[ht]
    \centering
    \includegraphics[scale=0.48]{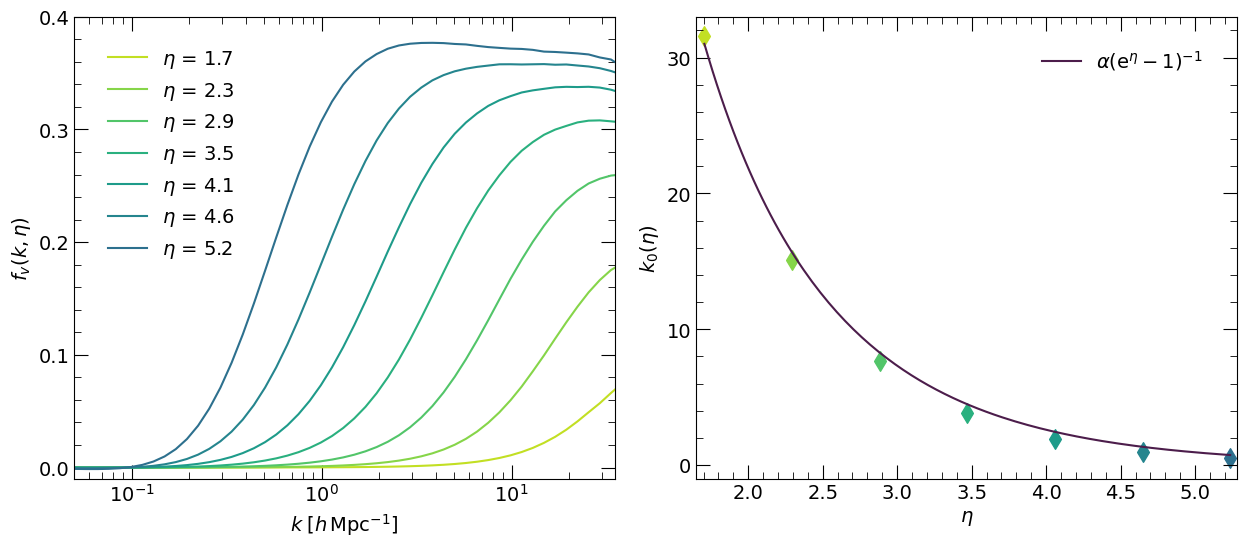}
    \caption{\textbf{Left panel:} $f_v$ for different times $\eta \in [0,6.75]$. One can observe nicely how the non-linear onset of the power spectrum shifts towards smaller $k$ over time, resulting in an interaction potential that becomes important on larger and larger scales as structure formation proceeds.
    \textbf{Right panel:}
    The time-dependence of the cut-off scale $k_0$.
    The coloured diamonds indicate the values of $k_0$ inferred by fitting the curves in the left panel at the corresponding times. The purple curve is a fit to the inferred value of $k_0$ with the indicated function.}
    \label{fig:apx}
\end{figure}

To determine the shape of the modified two-particle interaction potential \eqref{eq:modified two-particle potential} we need to find an approximation for $f_v(k,\eta)$.
In theory, this would require us to have full knowledge of the late time non-linear power spectrum.
However, we can approximate $f_v(k,\eta)$ by using the first order KPT result with a purely Newtonian two-particle interaction potential such that 
\begin{align}\label{eq:apx2 ansatz}
    P_\delta(k,\eta) \approx P^{(\mathrm{lin})}_\delta(k,\eta) + P^{(\mathrm{KPT},1)}_\delta(k,\eta)\big|_{f_v=0}.
\end{align}
Here, $P^{(\mathrm{KPT},1)}_\delta(k,\eta)\big|_{f_v=0}$ is the correction to the linear power spectrum due to first-order KPT, evaluated with the Newtonian potential ($f_v=0$).
We expect first-order KPT to be sufficient for this purpose, since we are mainly interested in the scale where the non-linear and the linear power spectra start deviating from each other. 
As is discussed in \autoref{sec:numerical results} corrections from higher order KPT introduce further power on smaller scales, while leaving the transition region between linear and non-linear growth untouched.
Inserting \eqref{eq:apx2 ansatz} into \eqref{eq:modified two-particle potential} we find
\begin{align*}
    f_v(k,\eta) \approx 1-\frac{1}{\sqrt{1+\Delta}}, \quad \mathrm{with}
    \qquad \Delta = \frac{P^{(\mathrm{KPT},1)}_\delta(k,\eta)\big|_{f_v=0}}{P^{(\mathrm{lin})}_\delta(k,\eta)},
\end{align*}
which is plotted in the left panel of \autoref{fig:apx} for different times.
Intuitively the modification that $f_v$ implies for the two-particle interaction potential can be well understood from \eqref{eq:mod Poisson_eq}.
There are two main regimes: 
(i) non-linear density fluctuations follow linear evolution, and (ii) non-linear density fluctuations greatly exceed linear growth.
According to \eqref{eq:mod Poisson_eq} linear density fluctuations source no interaction potential, whereas those fluctuations that greatly exceed linear growth result in an interaction potential that is nearly Newtonian (i.e.\ it fulfills the usual Poisson equation \eqref{eq:Poisson equation}).
It is well known \cite{BERNARDEAU20021} that large-scale (small $k$) structures evolve linearly, whereas on small scales (large $k$) density modes exceed linear growth.
As a consequence, we would expect the modified two-particle interaction potential to vanish for small $k$ and to attain its Newtonian strength as we go towards higher $k$.
This behaviour is indeed reproduced by the result from first-order KPT shown in the left panel of \autoref{fig:apx} .
Of course, we do not expect the amplitude of the first-order KPT power spectrum to match the full non-linear power spectrum on small scales and should therefore not take the amplitude of $f_v$ at large $k$ too seriously.
However, as argued above, the onset of the non-linear behaviour should be well represented already by first-order KPT results.
The functions shown in the left panel of \autoref{fig:apx} can be fitted by 
\begin{align*}
    f_v^{(\mathrm{fit})}(k,\eta) = A(\eta) \frac{k^2}{k^2 + k_0^2(\eta)},
\end{align*}
with an amplitude $A(\eta)$ and a time-dependent parameter $k_0(\eta)$ that characterizes the scale where structure growth becomes non-linear.
As mentioned above the amplitude $A$ is of little relevance to us, because it is unrealistically small in our first-order KPT calculation. 
We will therefore set it to 1 in our calculations, anticipating that the non-linear power spectrum greatly exceeds its linear counterpart on small scales.
In the right panel of \autoref{fig:apx} we plot the cut-off scale $k_0$ as a function of time, as well as a fit to this curve given by
\begin{align}\label{eq:Yukawa cutoff apx}
    k_0^{(\mathrm{fit})}(\eta) = \frac{a}{\ee^\eta-1}.
\end{align}
The best-fit parameter is $a=140\;h\,\mathrm{Mpc}^{-1}$.
Interestingly this roughly corresponds to the inverse length-scale set by the initial momentum variance $\sigma_p$, suggesting that non-linear growth sets in at wave-vectors $\vec{k}$ and times $\eta$ where
\begin{align*}
    \sigma_p \, |\vec{k}| \, g_{qp}^{(\mathcal{Z})}(\eta,0) \approx 1.
\end{align*}
We have thus shown that the modified two-particle interaction potential is approximately of Yukawa shape, characterized by a time-dependent cutoff which corresponds to the characteristic scale of the transition from linear to non-linear growth.
From the time-dependence of the Yukawa cutoff we can additionally infer that, as expected, non-linear growth sets in when the Gaussian damping factor starts to become important.

%%%%%%%%%%%%%%%%%%%%%%%%%%%%%%%%%%%%%%%%%%%%%%%%%%%%%%%
%%%%%%%%%%%%%%%%%%%%%%%%%%%%%%%%%%%%%%%%%%%%%%%%%%%%%%%
\section{Evaluation of second order diagrams} \label{apx:diagrams}
As shown in \autoref{fig:diagramsSecondOrder}, there are four distinct diagrams contributing to second-order KFT perturbation theory.
In the main text we have only shown the evaluation of one of these diagrams in terms of factors of the generating functional, namely of $\mathcal{D}^{(2)}_\mathrm{III}$, which is given by \eqref{eq:diagram 23} and \eqref{eq:generating functional second order}.
For completeness we summarize here the expressions for all four diagrams of second-order perturbation theory up to second order in initial power spectrum.
For each configuration the shift vectors in terms of the internal and external wave-vectors as well as the response field prefactors take different expressions. 
This in turn also changes the expressions of the resulting four-point density functions.

\subsection{Diagram I}
Using the definition of operator contractions in \eqref{eq:secondOrderContractions}, we find for diagram I:
\begin{align}\label{eq:diagram 21}
    \mathcal{D}^{(2)}_\mathrm{I}(1,2) = 
    \int \dd 2' \int \dd 1' \ v(2') v(1') b(1,-1') b(1,-2') G^{(0)}_{\rho\rho\rho\rho}(1-1'-2', 2, 1', 2').
\end{align}
The shift vectors are given by
\begin{align*}
    \vecL_{q_1} &= - \veck_1 + \veck'_1 + \veck'_2 \qquad  
    \vecL_{p_1} = - g_{qp}^{(\mathcal{Z})}(\eta_1,0) \veck_1 + g_{qp}^{(\mathcal{Z})}(\eta'_1,0) \veck'_1 + g_{qp}^{(\mathcal{Z})}(\eta'_2,0) \veck'_2
    \\
    \vecL_{q_2} &= - \veck_2 \qquad \qquad \qquad \ 
    \vecL_{p_2} = - g_{qp}^{(\mathcal{Z})}(\eta_1,0) \veck_2
    \\
    \vecL_{q_3} &= - \veck'_1 \qquad \qquad \qquad  \
    \vecL_{p_3} = - g_{qp}^{(\mathcal{Z})}(\eta'_1,0) \veck'_1
    \\
    \vecL_{q_4} &= - \veck'_2 \qquad \qquad \qquad \  
    \vecL_{p_4} = - g_{qp}^{(\mathcal{Z})}(\eta'_2,0) \veck'_2,
\end{align*}
and the factorization of the generating functional leads to the four-point function
\begin{multline}\label{eq:free cumulant 21}
    G^{(0)}_{\rho\rho\rho\rho}(1-1'-2', 2, 1', 2')=
    (2\pi)^3 \Bar{\rho}^{\,4} 
    \delta_D\big(\veck_1+ \veck_2 \big)
    \ee^{-{Q}_D(\tensorL_p)} 
    \Big[
    (2\pi)^3\delta_D(\vec{k}'_1 + \vec{k}'_2) \mathcal{P}_{21}(\vec{k}_1)\mathcal{P}_{43}(\vec{k}'_1)\\[2mm]
    +(2\pi)^3\delta_D(\vec{k}_1 - \vec{k}'_2) \mathcal{P}_{31}(\vec{k}'_1)\mathcal{P}_{42}(\vec{k}_1)
    +(2\pi)^3\delta_D(\vec{k}_1 - \vec{k}'_1) \mathcal{P}_{41}(\vec{k}'_2)\mathcal{P}_{32}(\vec{k}_1)\\[2mm]
    +(2\pi)^3\delta_D(\vec{k}_1 - \vec{k}'_1 - \vec{k}'_2) \mathcal{P}_{32}(\vec{k}'_1)\mathcal{P}_{42}(\vec{k}_1 - \vec{k}'_1) 
    +(2\pi)^3\delta_D(\vec{k}_1 - \vec{k}'_1 - \vec{k}'_2) \mathcal{P}_{32}(\vec{k}_1)\mathcal{P}_{43}(\vec{k}_1 - \vec{k}'_1)\\[2mm]
    +(2\pi)^3\delta_D(\vec{k}_1 - \vec{k}'_1 - \vec{k}'_2) \mathcal{P}_{42}(\vec{k}_1)\mathcal{P}_{43}(\vec{k}'_1)\Big].
\end{multline}

\subsection{Diagram II}
Using the definition of operator contractions in \eqref{eq:secondOrderContractions}, we find for diagram II:
\begin{align}\label{eq:diagram 22}
    \mathcal{D}^{(2)}_\mathrm{II}(1,2) = 
    \int \dd 2' \int \dd 1' \ v(2') v(1') b(1,-1') b(2,-2') G^{(0)}_{\rho\rho\rho\rho}(1-1', 2-2', 1', 2').
\end{align}
The shift vectors are given by
\begin{align*}
    \vecL_{q_1} &= - \veck_1 + \veck'_1 \qquad \qquad 
    \vecL_{p_1} = - g_{qp}^{(\mathcal{Z})}(\eta_1,0) \veck_1 + g_{qp}^{(\mathcal{Z})}(\eta'_1,0) \veck'_1
    \\
    \vecL_{q_2} &= - \veck_2 + \veck'_2  \qquad \qquad 
    \vecL_{p_2} = - g_{qp}^{(\mathcal{Z})}(\eta_1,0) \veck_2 + g_{qp}^{(\mathcal{Z})}(\eta'_2,0) \veck'_2
    \\
    \vecL_{q_3} &= - \veck'_1  \qquad \qquad \qquad  \, \vecL_{p_3} = - g_{qp}^{(\mathcal{Z})}(\eta'_1,0) \veck'_1
    \\
    \vecL_{q_4} &= - \veck'_2 \qquad \qquad \qquad \,  \vecL_{p_4} = - g_{qp}^{(\mathcal{Z})}(\eta'_2,0) \veck'_2.
\end{align*}
and the factorization of the generating functional leads to the four-point function
\begin{multline}\label{eq:free cumulant 22}
    G^{(0)}_{\rho\rho\rho\rho}(1-1', 2-2', 1', 2')=
    (2\pi)^3 \Bar{\rho}^{\,4} 
    \delta_D\big(\veck_1+ \veck_2 \big)
    \ee^{-{Q}_D(\tensorL_p)} 
    \Big[
    (2\pi)^3\delta_D(\vec{k}_1 - \vec{k}'_1) (2\pi)^3\delta_D(\vec{k}_1 + \vec{k}'_2) \mathcal{P}_{43}(\vec{k}_1)\\[2mm]
    +(2\pi)^3\delta_D(\vec{k}'_1 + \vec{k}'_2) \mathcal{P}_{21}(\vec{k}_1 - \vec{k}'_1)\mathcal{P}_{43}(\vec{k}'_1)
    +(2\pi)^3\delta_D(\vec{k}_1 + \vec{k}'_2) \mathcal{P}_{31}(\vec{k}'_1)\mathcal{P}_{41}(\vec{k}_1)\\[2mm]
    +(2\pi)^3\delta_D(\vec{k}_1 + \vec{k}'_2) \mathcal{P}_{31}(\vec{k}_1 - \vec{k}'_1)\mathcal{P}_{43}(\vec{k}_1)
    +(2\pi)^3\delta_D(\vec{k}_1 - \vec{k}'_1 + \vec{k}'_2) \mathcal{P}_{41}(\vec{k}_1 - \vec{k}'_1)\mathcal{P}_{32}(\vec{k}'_1)\\[2mm]
    +(2\pi)^3\delta_D(\vec{k}_1 + \vec{k}'_2) \mathcal{P}_{41}(\vec{k}_1 - \vec{k}'_1)\mathcal{P}_{43}(\vec{k}'_1)
    +(2\pi)^3\delta_D(\vec{k}_1 - \vec{k}'_1) \mathcal{P}_{32}(\vec{k}_1)\mathcal{P}_{42}(\vec{k}'_2)\\[2mm]
    +(2\pi)^3\delta_D(\vec{k}_1 - \vec{k}'_1) \mathcal{P}_{32}(\vec{k}_1 + \vec{k}'_2)\mathcal{P}_{43}(\vec{k}'_2) 
    +(2\pi)^3\delta_D(\vec{k}_1 - \vec{k}'_1) \mathcal{P}_{42}(\vec{k}_1 + \vec{k}'_2)\mathcal{P}_{43}(\vec{k}_1)\Big].
\end{multline}

\subsection{Diagram III}
Using the definition of operator contractions in \eqref{eq:secondOrderContractions}, we find for diagram III:
\begin{align}\label{eq:diagram 23 apx}
    \mathcal{D}^{(2)}_\mathrm{III}(1,2) = 
    \int \dd 2' \int \dd 1' \ v(2') v(1') b(1,-1') b(1',-2') G^{(0)}_{\rho\rho\rho\rho}(1-1', 2, 1'-2', 2').
\end{align}
The shift vectors are given by
\begin{align*}
    \vecL_{q_1} &= - \veck_1 + \veck'_1 \qquad \qquad \vecL_{p_1} = - g_{qp}^{(\mathcal{Z})}(\eta_1,0) \veck_1 + g_{qp}^{(\mathcal{Z})}(\eta'_1,0) \veck'_1
    \\
    \vecL_{q_2} &= - \veck_2 \qquad \qquad \qquad \, \vecL_{p_2} = - g_{qp}^{(\mathcal{Z})}(\eta_1,0) \veck_2
    \\
    \vecL_{q_3} &= - \veck'_1 + \veck'_2 \qquad \qquad   \vecL_{p_3} = - g_{qp}^{(\mathcal{Z})}(\eta'_1,0) \veck'_1 + g_{qp}^{(\mathcal{Z})}(\eta'_2,0) \veck'_2
    \\
    \vecL_{q_4} &= - \veck'_2 \qquad \qquad \qquad \,  \vecL_{p_4} = - g_{qp}^{(\mathcal{Z})}(\eta'_2,0) \veck'_2.
\end{align*}
and the factorization of the generating functional leads to the four-point function
\begin{multline}\label{eq:free cumulant 23}
    G^{(0)}_{\rho\rho\rho\rho}(1-1', 2, 1'-2', 2')=
    (2\pi)^3 \Bar{\rho}^{\,4} 
    \delta_D\big(\veck_1+ \veck_2 \big)
    \ee^{-{Q}_D(\tensorL_p)} 
    \Big[
    (2\pi)^3\delta_D(\vec{k}_1 - \vec{k}'_1) (2\pi)^3\delta_D(\vec{k}_1 - \vec{k}'_2) \mathcal{P}_{42}(\vec{k}_1)\\[2mm]
    +(2\pi)^3\delta_D(\vec{k}'_1 - \vec{k}'_2) \mathcal{P}_{21}(\vec{k}_1)\mathcal{P}_{41}(\vec{k}'_1)
    +(2\pi)^3\delta_D(\vec{k}'_1 - \vec{k}'_2) \mathcal{P}_{21}(\vec{k}_1 - \vec{k}'_1)\mathcal{P}_{42}(\vec{k}'_1)\\[2mm]
    +(2\pi)^3\delta_D(\vec{k}_1 - \vec{k}'_2) \mathcal{P}_{31}(\vec{k}_1 - \vec{k}'_1)\mathcal{P}_{42}(\vec{k}_1)
    +(2\pi)^3\delta_D(\vec{k}_1 - \vec{k}'_1 + \vec{k}'_2) \mathcal{P}_{41}(\vec{k}_1 - \vec{k}'_1)\mathcal{P}_{32}(\vec{k}_1)\\[2mm]
    +(2\pi)^3\delta_D(\vec{k}'_1 - \vec{k}'_2) \mathcal{P}_{41}(\vec{k}_1 - \vec{k}'_1)\mathcal{P}_{42}(\vec{k}_1)
    +(2\pi)^3\delta_D(\vec{k}_1 - \vec{k}'_1) \mathcal{P}_{32}(\vec{k}_1 - \vec{k}'_2)\mathcal{P}_{42}(\vec{k}'_2)\\[2mm]
    +(2\pi)^3\delta_D(\vec{k}_1 - \vec{k}'_1) \mathcal{P}_{32}(\vec{k}_1)\mathcal{P}_{43}(\vec{k}'_2)
    +(2\pi)^3\delta_D(\vec{k}_1 - \vec{k}'_1) \mathcal{P}_{42}(\vec{k}_1)\mathcal{P}_{43}(\vec{k}_1 - \vec{k}'_2) +\mathcal{O}(\curlyP^3)\Big].
\end{multline}

\subsection{Diagram IV}
Using the definition of operator contractions in \eqref{eq:secondOrderContractions}, we find for diagram IV:
\begin{align}\label{eq:diagram 24}
    \mathcal{D}^{(2)}_\mathrm{IV}(1,2) = 
    \int \dd 2' \int \dd 1' \ v(2') v(1') b(1,-1') b(-1',-2') G^{(0)}_{\rho\rho\rho\rho}(1-1'-2', 2, 1', 2').
\end{align}
The shift vectors and the four-point function of this diagram are identical to those of diagram I. 
The only difference between diagrams I and IV thus lies in the response field prefactors.

\bibliographystyle{ieeetr}
\bibliography{biblio}

\end{document}